\documentclass[a4paper,11pt]{article}
\pdfoutput=1 

\usepackage{jcappub} 

\usepackage[T1]{fontenc} 
\usepackage{subcaption}
\usepackage{pgfplots}
\usepackage{float}
\usepackage[export]{adjustbox}
\usepackage{natbib}
\usepackage{color}

\usepackage[normalem]{ulem}

\renewcommand{\v}[1]{\boldsymbol{#1}}
\def\vtheta{\v{\theta}}
\def\valpha{\v{\alpha}}
\def\vell{\v{\ell}}
\def\vx{\v{x}}
\def\vk{\v{k}}
\def\eps{\epsilon}
\newcommand{\bq}{\begin{eqnarray}}
\newcommand{\eq}{\end{eqnarray}}

\title{Cosmology from the integrated shear 3-point correlation function: \vspace{1mm} \newline \Large simulated likelihood analyses with machine-learning emulators}

\author[a,b]{Zhengyangguang Gong,}
\author[a,b]{Anik Halder,}
\author[c,d]{Alexandre Barreira,}
\author[a,b]{Stella Seitz,}
\author[a]{Oliver Friedrich}

\affiliation[a]{Universit\"{a}ts-Sternwarte, Fakult\"{a}t f\"{u}r Physik, Ludwig-Maximilians Universit\"{a}t M\"{u}nchen,\\Scheinerstra{\ss}e 1, 81679 M\"{u}nchen, Germany}
\affiliation[b]{Max Planck Institute for Extraterrestrial Physics, Giessenbachstra{\ss}e 1, 85748 Garching, Germany}
\affiliation[c]{Excellence Cluster ORIGINS, Boltzmannstra{\ss}e 2, 85748 Garching, Germany}
\affiliation[d]{Ludwig-Maximilians-Universit\"{a}t, Schellingstra{\ss}e 4, 80799 M\"{u}nchen, Germany}

\emailAdd{lgong@usm.lmu.de}
\emailAdd{ahalder@usm.lmu.de}
\emailAdd{alex.barreira@origins-cluster.de}
\emailAdd{stella@usm.lmu.de}
\emailAdd{oliver.friedrich@physik.uni-muenchen.de}

\abstract{The integrated shear 3-point correlation function $\zeta_{\pm}$ measures the correlation between the local shear 2-point function $\xi_{\pm}$ and the 1-point shear aperture mass in patches of the sky. Unlike other higher-order statistics, $\zeta_{\pm}$ can be efficiently measured from cosmic shear data, and it admits accurate theory predictions on a wide range of scales as a function of cosmological and baryonic feedback parameters. Here, we develop and test a likelihood analysis pipeline for cosmological constraints using $\zeta_{\pm}$. We incorporate treatment of systematic effects from photometric redshift uncertainties, shear calibration bias and galaxy intrinsic alignments. We also develop an accurate neural-network emulator for fast theory predictions in MCMC parameter inference analyses. We test our pipeline using realistic cosmic shear maps based on $N$-body simulations with a DES Y3-like footprint, mask and source tomographic bins, finding unbiased parameter constraints. Relative to $\xi_{\pm}$-only, adding $\zeta_{\pm}$ can lead to $\approx 10-25\%$ improvements on the constraints of parameters like $A_s$ (or $\sigma_8$) and $w_0$. We find no evidence in $\xi_{\pm} + \zeta_{\pm}$ constraints of a significant mitigation of the impact of systematics. We also investigate the impact of the size of the apertures where $\zeta_{\pm}$ is measured, and of the strategy to estimate the covariance matrix ($N$-body vs.~lognormal). Our analysis solidifies the strong potential of the $\zeta_{\pm}$ statistic and puts forward a pipeline that can be readily used to improve cosmological constraints using real cosmic shear data.}

\begin{document}
\maketitle
\flushbottom

\section{Introduction}
\label{sec:intro}

The weak gravitational lensing effect is the bending of the light of background source galaxies by foreground gravitational potentials \cite{2001PhR...340..291B, 2006glsw.conf.....M}. This induces a coherent distortion pattern in the observed shape of the background galaxies that is called the {\it cosmic shear field}. The statistics of this field depend on the three-dimensional large-scale structure, hence cosmic shear studies offer a powerful way to address key questions in cosmology such as the structure formation history, the nature of dark energy and dark matter, and the laws of gravity on large scales. Indeed, cosmic shear is one of the most active research areas in large-scale structure today: the DES \cite{2022PhRvD.105b3520A}, KiDS \cite{2021A&A...645A.104A} and HSC-SSP \cite{2020PASJ...72...16H} surveys have recently presented cosmological constraints from their cosmic shear data, and more accurate and bigger data sets will be available soon with missions like Euclid \cite{2020A&A...635A.139E}, Vera Rubin's LSST \cite{2012arXiv1211.0310L} and Nancy Roman \cite{2023MNRAS.519.4241Y}. 

\vspace{1mm}
The majority of cosmic shear analyses are based on the shear 2-point correlation function (2PCF), or its Fourier counterpart the lensing power spectrum. These statistics completely characterize the information content of Gaussian random fields, which our Universe was close to at the earliest stages of its evolution, as well as today on sufficiently large-scales. At late times, however, the evolution of matter density fluctuations becomes nonlinear on small scales, inducing non-Gaussian features in the cosmic shear field that cannot be described by 2PCF alone. Higher-order statistics are thus needed to access the non-Gaussian information.

\vspace{1mm}
The shear 3-point correlation function (3PCF; or its Fourier counterpart the lensing bispectrum) is the natural first step beyond the 2PCF \cite{2003A&A...397..809S, 2004MNRAS.348..897T, 2005A&A...431....9S, 2005PhRvD..72h3001D, 2013MNRAS.429..344K, 2013PhRvD..87l3538S}. However, being a more complicated statistic, it is more challenging to measure observationally, as well as to predict theoretically. Concretely, compared to the 2PCF which depends only on the distance between two points in the survey footprint, the 3PCF is a function of the size and shape of triangles connecting three points, which requires more demanding estimators. Additionally, theoretical predictions require accurate prescriptions for the nonlinear matter bispectrum, which despite recent progress \cite{2016MNRAS.455.2945M, 2020ApJ...895..113T}, are still not as developed as the matter power spectrum that enters the shear 2PCF. Further complications arise by the need to account for baryonic feedback effects, as well as systematics effects such as photometric redshift uncertainties, shear multiplicative bias and galaxy intrinsic alignments (IA). This helps explain why existing real-data constraints using higher-order shear information are based not on the full 3-point correlation function, but on other statistics including aperture moments \cite{2011MNRAS.410..143S, 2014MNRAS.441.2725F, Barthelemy2020, 2022PhRvD.105j3537S, 2022arXiv220811686H}, lensing peaks \cite{Kacprzak2016MNRAS, Harnois-Deraps2021, zurcher2022}, density-split statistics \cite{Friedrich_2018, gruen_2018, Burger_2020, burger2021revised, burger2022} and persistent homology of cosmic shear \cite{2021A&A...648A..74H, 2022A&A...667A.125H}. The shear 3PCF was recently measured using DES Year 3 (Y3) data \cite{2022PhRvD.105j3537S}, although only in patches over the survey and not over the whole footprint as that would be too computationally demanding.

\vspace{1mm}
In this paper, we focus on a particular kind of shear 3PCF called the \textit{integrated shear 3-point correlation function} \cite{2021MNRAS.506.2780H}. This statistic corresponds to the correlation between the shear 2PCF measured in patches of the sky with the 1-point shear aperture mass in those patches.\footnote{See also Refs.~\cite{2014JCAP...05..048C, Chiang_2015} for earlier applications of the same idea in the context of the three-dimensional galaxy distribution, and Refs.~\cite{2017JCAP...02..010M, 2021JCAP...06..055J, 2023PhRvD.107d3516M} for studies of the Fourier counterpart of the integrated shear 3PCF.} Physically, this statistic describes the modulation of the local shear 2PCF by long-wavelength features in the cosmic shear field. The integrated shear 3PCF enjoys two key advantages relative to other higher-order shear statistics. The first is that it is straightforward to measure from the data as it requires only conventional and well-tested shear 2PCF estimators. The second is that, as shown in Ref.~\cite{2022MNRAS.515.4639H}, this statistic is sensitive to the squeezed matter bispectrum that can be evaluated accurately in the nonlinear regime using the response approach to perturbation theory \cite{2017JCAP...06..053B}. Importantly, the response approach allows to account for the impact of baryonic feedback on small scales, which is crucial to design scale cuts and/or marginalize over these uncertainties in real data analyses.

\vspace{1mm}
Our goal here is to develop and test a likelihood analysis pipeline to reliably extract cosmology from real cosmic shear data using the integrated shear 3PCF. Concretely, we incorporate the impact of baryonic feedback (as in Ref.~\cite{2022MNRAS.515.4639H}), as well as of photometric redshift uncertainties, shear multiplicative bias and galaxy IA. We also develop a neural-network (NN) emulator for the theory model to enable fast theory predictions in Monte-Carlo Markov Chain (MCMC) parameter inference analyses. We test our analysis pipeline on simulated cosmic shear maps with DES Y3-like survey footprints and source galaxy redshift distributions. We study in particular (i) the ability of the theory model to return unbiased parameter constraints\footnote{Throughout the paper we loosely use the term ``unbiased constraints'' to mean that the $68\%$ posterior credible intervals encompass the true model parameter values.}, (ii) the impact of the size of the aperture where the integrated shear 3PCF is measured, (iii) the ability of combined 2PCF and 3PCF analyses to mitigate the impact of systematic uncertainties, and (iv) the impact of different data vector covariance estimates.

\vspace{1mm}
In terms of constraining power, we find that the integrated shear 3PCF leads to improvements of $\approx 10-25\%$ on the constraints of parameters like the amplitude of primordial density fluctuations $A_s$ (or equivalently $\sigma_8)$ or the dark energy equation of state parameter $w_0$. This is consistent with the previous findings of Refs.~\cite{2021MNRAS.506.2780H, 2022MNRAS.515.4639H} based on idealized Fisher matrix forecasts, but now in the context of realistically simulated MCMC likelihood analyses. Our results thus strongly motivate as next steps exploring the power of this statistic to improve cosmological constraints using real cosmic shear data.

\vspace{1mm}
This paper is structured as follows: In Sec.~\ref{sec:theory} we review the theoretical formalism behind the integrated shear 3PCF and describe how we incorporate lensing systematic effects. In Sec.~\ref{sec:cov} we describe the construction of our DES Y3-like cosmic shear maps, as well as the measurements of the shear 2PCF, integrated 3PCF and their (cross) covariance matrices. We describe and discuss the performance of our NN emulator of the theory predictions for fast MCMC likelihood analyses in Sec.~\ref{sec:emulation}. Our main numerical results are shown in Sec.~\ref{sec:results}. We summarize and conclude in Sec.~\ref{sec:conclusion}. Appendix \ref{appendix:NLA} describes our modelling of the galaxy IA.

\section{Theoretical formalism}
\label{sec:theory}
In this section we describe the theory behind the integrated shear 3PCF.  We begin with a recap of the model of Refs.~\cite{2021MNRAS.506.2780H, 2022MNRAS.515.4639H}, and then discuss how we incorporate lensing systematics.

\subsection{Integrated shear 3-point correlation function}
\label{sec:theory_recap}

The integrated shear 3PCF, $\zeta_{\pm, ijk}(\boldsymbol{\alpha})$, is defined as
\begin{equation}
\label{eq:i3pcf_ensemble_average}
    \zeta_{\pm, ijk}(\boldsymbol{\alpha}) \equiv \big<M_{\text{ap},i}(\boldsymbol{\theta}_C)\hat{\xi}_{\pm, jk}(\boldsymbol{\alpha}; \boldsymbol{\theta}_C)\big> \ ,
\end{equation}
where $M_{\text{ap},i}(\boldsymbol{\theta}_C)$ is the 1-point aperture mass statistic measured on a patch of the survey centered at angular position $\boldsymbol{\theta}_C$, and $\hat{\xi}_{\pm, jk}(\boldsymbol{\alpha}; \boldsymbol{\theta}_C)$ is the shear 2PCF measured on the same patch of the sky; $\boldsymbol{\alpha}$ describes angular separations. The angle brackets denote ensemble average (or in practice, averaging over all positions $\boldsymbol{\theta}_C$) and the subscripts $i,j,k$ denote tomographic source bins, i.e.~$\hat{\xi}_{\pm, jk}$ is the 2PCF of the shear fields from galaxy shape measurements at the redshift bins $j$ and $k$. This equation makes apparent the interpretation of the shear 3PCF as describing the spatial modulation of the local 2PCF by the local shear mass aperture, which describes larger-scale features in the shear field.

\vspace{1mm}
The aperture mass $M_{\text{ap}}(\boldsymbol{\theta}_C)$ is defined as \cite{2006glsw.conf.....M, 1996MNRAS.283..837S}
\begin{equation}
\label{eq:aperture_mass_kappa}
    M_{\text{ap}}(\boldsymbol{\theta}_C) = \int \text{d}^2\boldsymbol{\theta} \ \kappa(\boldsymbol{\theta})U(\boldsymbol{\theta}_C - \boldsymbol{\theta}) \ ,
\end{equation}
where $\kappa(\boldsymbol{\theta})$ is the lensing convergence field, and $U$ is an azimuthally symmetric filter function with angular size $\theta_{\text{ap}}$. The convergence field is not directly observable, but if $U$ is a compensated filter satisfying $\int \text{d}^2\boldsymbol{\theta} \ U(\boldsymbol{\theta}_C - \boldsymbol{\theta}) = 0$, then $M_{\text{ap}}(\boldsymbol{\theta}_C)$ can be expressed as
\begin{equation}
\label{eq:aperture_mass_Q}
    M_{\text{ap}}(\boldsymbol{\theta}_C) = \int \text{d}^2\boldsymbol{\theta} \ \gamma_{\text{t}}(\boldsymbol{\theta}, \phi_{\boldsymbol{\theta}_C - \boldsymbol{\theta}})Q(\boldsymbol{\theta}_C - \boldsymbol{\theta}) \ ,
\end{equation}
where $\gamma_{\text{t}}$ is the tangential component of the shear field (which is directly observable), $\phi_{\boldsymbol{\theta}_C - \boldsymbol{\theta}}$ is the polar angle of the angular separation between $\boldsymbol{\theta}_C$ and $\boldsymbol{\theta}$, and $Q$ is a filter function related to $U$. As in previous works, we adopt the following form for $U$ and $Q$ \cite{2002ApJ...568...20C}
\begin{align}
    \label{eq:crittenden_compensated_filter_U}
    U(\theta) &= \frac{1}{2\pi\theta^2_{\text{ap}}}\left(1 - \frac{\theta^2}{2\theta^2_{\text{ap}}}\right)\text{exp}\left(-\frac{\theta^2}{2\theta^2_{\text{ap}}}\right)\ , \\
    \label{eq:crittenden_compensated_filter_Q}
    Q(\theta) &= \frac{\theta^2}{4\pi\theta^2_{\text{ap}}}\text{exp}\left(-\frac{\theta^2}{2\theta^2_{\text{ap}}}\right) \ ;
\end{align}
note the filters depend only on the magnitude of the arguments because of the azimuthal symmetry. The Fourier transform of $U$, which appears in equations below, is given by
\begin{equation}
    \label{eq:crittenden_compensated_filter_U_fourier}
    U(\ell) = \int \text{d}^2\boldsymbol{\theta} \ U(\theta) e^{-i\boldsymbol{\ell}\cdot\boldsymbol{\theta}} = \frac{\ell^2\theta^2_{\text{ap}}}{2}\text{exp}\left(-\frac{\ell^2\theta^2_{\text{ap}}}{2}\right) \ ,
\end{equation}
where $\boldsymbol{\ell}$ is a two-dimensional wavevector on the sky (we assume the flat-sky approximation).

\vspace{1mm}
The other term in Eq.~\eqref{eq:i3pcf_ensemble_average}, $\hat{\xi}_{\pm}(\boldsymbol{\alpha};\boldsymbol{\theta}_C)$, is the 2PCF of the {\it windowed} shear field $\gamma(\boldsymbol{\theta};\boldsymbol{\theta}_C) \equiv \gamma(\boldsymbol{\theta})W(\boldsymbol{\theta}_C-\boldsymbol{\theta})$, where the window function $W$ is a top-hat of size $\theta_{\rm T}$ at position $\boldsymbol{\theta}_C$. The two 2PCFs are defined as
\begin{equation} \label{eq:position_dependent_2pt_function_2D_field_xipm}
\begin{split}
    \hat{\xi}_{+}(\boldsymbol{\alpha};\boldsymbol{\theta}_C) & \equiv \frac{1}{A_{\mathrm{2pt}}(\boldsymbol{\alpha})} \int \mathrm{d}^2 \boldsymbol{\theta} \; \gamma(\boldsymbol{\theta};\boldsymbol{\theta}_C) \gamma^*(\boldsymbol{\theta}+\boldsymbol{\alpha};\boldsymbol{\theta}_C)  \,  \\
    \hat{\xi}_{-}(\boldsymbol{\alpha};\boldsymbol{\theta}_C) & \equiv \frac{1}{A_{\mathrm{2pt}}(\boldsymbol{\alpha})} \int \mathrm{d}^2 \boldsymbol{\theta} \; \gamma(\boldsymbol{\theta};\boldsymbol{\theta}_C) \gamma(\boldsymbol{\theta}+\boldsymbol{\alpha};\boldsymbol{\theta}_C) e^{-4i\phi_{\boldsymbol{\alpha}}}, \\
\end{split}
\end{equation}
where $^*$ denotes complex conjugation, $\phi_{\boldsymbol{\alpha}}$ is the polar angle of $\boldsymbol{\alpha}$, and $A_{\mathrm{2pt}}(\boldsymbol{\alpha}) \equiv \int \mathrm{d}^2 \boldsymbol{\theta} \;  W(\boldsymbol{\theta}_C-\boldsymbol{\theta}) W(\boldsymbol{\theta}_C-\boldsymbol{\theta}-\boldsymbol{\alpha})$. The Fourier transform of $W$ appears in equations below, and is given by
\begin{equation} \label{eq:tophat_window_function_unnormalised}
\begin{split}
    W(\boldsymbol{l}) = W(l) & = 2\pi \theta_{\mathrm{T}}^2 \; \frac{J_1(l\theta_{\mathrm{T}})}{l\theta_{\mathrm{T}}} ,
\end{split}
\end{equation}
where $J_{n}$ is the $n$th-order Bessel function of the first kind.

\vspace{1mm}
Skipping the details of the derivation \cite{2021MNRAS.506.2780H}, the two 3PCF in Eq.~\eqref{eq:i3pcf_ensemble_average} can be written as
 \begin{align}
    \label{eq:shear_i3pcf_plus}
    \zeta_{+, ijk}(\boldsymbol{\alpha}) &= \frac{1}{A_{\text{2pt}}(\alpha)}\int \frac{\text{d}\ell \ell}{2\pi} \mathcal{B}^{\rm 2D}_{+, ijk}(\boldsymbol{\ell}) J_0(\ell\alpha)\ , \\
    \label{eq:shear_i3pcf_minus}
    \zeta_{-, ijk}(\boldsymbol{\alpha}) &= \frac{1}{A_{\text{2pt}}(\alpha)}\int \frac{\text{d}\ell \ell}{2\pi} \mathcal{B}^{\rm 2D}_{-, ijk}(\boldsymbol{\ell}) J_4(\ell\alpha) \ ,
\end{align}
where $\mathcal{B}^{\rm 2D}_{\pm}$ is called the integrated lensing bispectrum, and it is given by (in the Limber approximation)
\begin{equation}
    \label{eq:integrated_bispectrum_2d}
    \begin{split}
    \mathcal{B}^{\rm 2D}_{\pm, ijk}(\boldsymbol{\ell}) &= \int \text{d}\chi \frac{q^i(\chi)q^j(\chi)q^k(\chi)}{\chi^4}\int \frac{\text{d}^2\boldsymbol{\ell}_1}{(2\pi)^2} \int\frac{\text{d}^2\boldsymbol{\ell}_2}{(2\pi)^2} B^{\rm 3D}_{\delta}\left(\frac{\boldsymbol{\ell}_1}{\chi}, \frac{\boldsymbol{\ell}_2}{\chi}, \frac{-\boldsymbol{\ell}_1-\boldsymbol{\ell}_2}{\chi}, \chi\right) \\
    &\times e^{2i(\phi_2 \mp \phi_{-1-2})}U(\boldsymbol{\ell}_1)W(\boldsymbol{\ell}_2+\boldsymbol{\ell})W(-\boldsymbol{\ell}_1-\boldsymbol{\ell}_2-\boldsymbol{\ell}) \ .
    \end{split}
\end{equation}
In this equation, $B^{3D}_{\delta}$ is the 3-dimensional matter bispectrum (discussed below), $\phi_2$ is the polar angle of $\boldsymbol{\ell}_2$, $\phi_{-1-2}$ is the polar angle of $-\boldsymbol{\ell}_1-\boldsymbol{\ell}_2$, and $q(\chi)$ is the lensing kernel 
\begin{equation}
    \label{eq:lensing_kernel_q}
    q^i(\chi) = \frac{3H_0^2\Omega_{m}}{2c^2}\frac{\chi}{a(\chi)}\int_{\chi} \text{d}\chi^{\prime} n_s^i(\chi^{\prime}) \frac{\chi^{\prime} - \chi}{\chi^{\prime}} \ ,
\end{equation}
where $n_s^i(\chi)$ is the galaxy source number density distribution for the redshift tomographic bin $i$, $\chi$ denotes comoving distances, $H_0$ is the Hubble parameter, $\Omega_{m}$ is the cosmic matter density parameter today, $c$ is the speed of light and $a(\chi)$ is the scale factor; note that throughout the paper we always assume spatially flat cosmologies.

\vspace{1mm}
In our results, we will consider also the \textit{global} shear 2PCF, which can be evaluated as
\begin{align}
    \label{eq:shear_2pcf_plus}
    \xi_{+, ij}({\alpha}) &= \int \frac{\text{d}\ell \ell}{2\pi} P_{\kappa, ij} ({\ell}) J_0(\ell\alpha)\ , \\
    \label{eq:shear_2pcf_minus}
    \xi_{-, ij}({\alpha}) &= \int \frac{\text{d}\ell \ell}{2\pi} P_{\kappa, ij} ({\ell}) J_4(\ell\alpha) \ ,
\end{align}
where $P_{\kappa, ij}$ is the convergence power spectrum given by (in the Limber approximation)
\begin{equation}
	\label{eq:kappa_power_spectrum}
	P_{\kappa, ij}({\ell}) = \int \text{d}\chi \frac{q^i(\chi)q^j(\chi)}{\chi^2}P^{3D}_{\delta} \left({k}=\frac{{\ell}}{\chi},\chi \right) \ ,
\end{equation}
with $P^{3D}_{\delta}$ the three-dimensional matter power spectrum.


\subsection{The three-dimensional matter bispectrum model}
\label{sec:theory_recap_3dbispectrum}

A key ingredient to evaluate $\zeta_{\pm}$ is the three-dimensional matter bispectrum $B^{\rm 3D}_{\delta}$ in Eq.~\eqref{eq:integrated_bispectrum_2d}, which we evaluate following Ref.~\cite{2022MNRAS.515.4639H} as
\begin{equation}
\label{eq:3d_bispectrum_rf_gm}
B^{\rm 3D}_{\delta}(\boldsymbol{k_1}, \boldsymbol{k_2}, \boldsymbol{k_3}, \chi) =
	\begin{cases}
		B^{\rm 3D}_{\delta, \rm RF} \ , & f_{\rm sq} \geq f_{\rm sq}^{\rm thr} \Longrightarrow {\rm squeezed}  \\
		B^{\rm 3D}_{\delta, \rm GM} \ , & {\rm otherwise}
	\end{cases} \ ,
\end{equation}
where $B^{\rm 3D}_{\delta, \rm RF}$ is the bispectrum expression of the response function approach valid for squeezed configurations, and $B^{\rm 3D}_{\delta, \rm GM}$ is the bispectrum fitting formula of Ref.~\cite{2012JCAP...02..047G}. The parameter $f_{\rm sq}$ is defined as $f_{\rm sq} = k_m/k_s$, with $k_s$ ($k_m$) the smallest (intermediate) of the amplitudes of the three modes $\boldsymbol{k_i}$. As explained in Ref.~\cite{2022MNRAS.515.4639H}, this equation guarantees that the response function branch correctly evaluates the squeezed matter bispectrum configurations in the nonlinear regime, which determine the value of $\zeta_{\pm}$ on small angular scales. The value of $f_{\rm sq}^{\rm thr}$ is the threshold that defines whether a given bispectrum configuration is dubbed as squeezed or not. Ref.~\cite{2022MNRAS.515.4639H} found that a range of values around $f_{\rm sq}^{\rm thr} \approx 7$ yield good fits to simulation measurements; in this paper we adopt $f_{\rm sq}^{\rm thr} = 7$.

The response function branch in Eq.~(\ref{eq:3d_bispectrum_rf_gm}) is given by
\begin{align}
\label{eq:3d_bispectrum_rf}
B^{3D}_{\delta, \text{RF}}(\boldsymbol{k}_1, \boldsymbol{k}_2, \boldsymbol{k}_3, z) &= \left[R_1(k_h, z) + \left(\mu^2_{\boldsymbol{k}_h, \boldsymbol{k}_s} - \frac{1}{3}\right)R_K(k_h, z)\right] P^{3D}_{\delta}(k_h,z)P^{3D}_{\delta, L}(k_s,z) \ ,
\end{align} 
where $\boldsymbol{k}_h$ denotes the mode $\boldsymbol{k}_i$ with the highest magnitude, $\mu_{\boldsymbol{k}_i, \boldsymbol{k}_j}$ is the cosine of the angle between $\boldsymbol{k}_i$ and $\boldsymbol{k}_j$, $P^{3D}_{\delta, L}$ is the three-dimensional linear matter power spectrum and $R_1(k, z)$ and $R_K(k, z)$ are the first-order response functions of the matter power spectrum to large-scale density and tidal fields:
\begin{equation}  \label{eq:R_1}
    R_1(k,z) = 1 - \frac{1}{3} \frac{\mathrm{d \ln} P_{\delta}^{\mathrm{3D}}(k,z)}{\mathrm{d \ln} k} + G_1(k,z), \\
\end{equation}
\begin{equation} \label{eq:R_K}
    R_K(k,z) = G_K(k,z) - \frac{\mathrm{d \ln} P_{\delta}^{\mathrm{3D}}(k,z)}{\mathrm{d \ln} k}.
\end{equation}
In these expressions, $G_1$ and $G_K$ are the so-called {\it growth-only} response functions, which can be measured in the nonlinear regime of structure formation using separate universe simulations. Just as in Ref.~\cite{2022MNRAS.515.4639H}, we use the results of Ref.~\cite{Wagner2015} for $G_1$ and Ref.~\cite{Schmidt2018} for $G_K$.

The GM branch is in turn given by
\begin{align}
\label{eq:3d_bispectrum_gm}
B^{3D}_{\delta, \text{GM}}(\boldsymbol{k}_1, \boldsymbol{k}_2, \boldsymbol{k}_3, z) &= 2F^{\text{eff}}_2(\boldsymbol{k}_1, \boldsymbol{k}_2, z)P^{3D}_{\delta}(k_1,z)P^{3D}_{\delta}(k_2,z) + \text{cyclic permutations} \ ,
\end{align} 
where $F^{\text{eff}}_2(\boldsymbol{k}_1, \boldsymbol{k}_2, z)$ is a modified version of the 2-point mode coupling kernel with free functions calibrated against $N$-body simulations \cite{2012JCAP...02..047G}.

In this paper, we evaluate the nonlinear matter power spectrum using the \texttt{HMcode} \cite{2015MNRAS.454.1958M} implementation inside the publicly available Boltzmann code \texttt{CLASS} \cite{2011JCAP...07..034B}; to model the impact of baryonic feedback effects, we adopt the single parameter $c_{\rm min}$ parametrization, where $c_{\rm min}$ roughly describes the strength of feedback by active galactic nuclei (AGN). As discussed in Refs.~\cite{2019MNRAS.488.2079B, 2020MNRAS.498.2887F}, mode-coupling terms like $F^{\text{eff}}_2$, $G_1$ and $G_K$ are expected to be very weakly dependent on baryonic physics. This way, the impact of baryonic effects on the bispectrum is trivially propagated by that on the power spectrum; note that in practice the baryonic effects impact only the response function branch in Eq.~(\ref{eq:3d_bispectrum_rf_gm}), since the GM branch contributes only on large scales \cite{2022MNRAS.515.4639H} where baryonic effects have a negligible role.


\subsection{Systematic error effects}
\label{sec:systematics}

Reference \cite{2022MNRAS.515.4639H} has shown how to include the impact of baryonic feedback effects on $\zeta_{\pm}$, which are one of the main non-cosmological contaminants in cosmic shear analyses. In this subsection we describe how we take into account a series of other important systematic effects, namely photometric redshift uncertainties, multiplicative shear bias and galaxy IA.

Photometric redshift (photo-$z$) uncertainties have a direct impact on the galaxy source redshift distribution. Here, we follow a strategy commonly adopted in real-data analyses and parametrize their effect through a single shift parameter $\Delta z$ defined as
\begin{equation}
    \label{eq:photometric_uncertainty_model}
    n^i_s(z) = \hat{n}^i_s(z + \Delta z^i) \ ,
\end{equation}
where $\hat{n}^i_s$ is the default estimate for the galaxy source redshift bin $i$. This simple way to account for photo-$z$ uncertainties was found sufficient at the statistical power of DES-Y3 analyses (see Figure 10 in Ref.~\cite{2022PhRvD.105b3514A}).

Again, as common in the literature, we model biases from the shear measurement pipeline with multiplicative factors $1+m_i$ for each tomographic bin $i$. In practice, this implies the following transformations of $\xi_{\pm}$ and $\zeta_{\pm}$,
\begin{align}
    \label{eq:shear_2pcf_multiplicative_bias_model}
    \xi_{\pm, ij}(\alpha) &\longrightarrow (1+m_i)(1+m_j)\xi_{\pm, ij}(\alpha) \ , \\
    \label{eq:shear_i3pcf_multiplicative_bias_model}
    \zeta_{\pm, ijk}(\alpha) &\longrightarrow (1+m_i)(1+m_j)(1+m_k)\zeta_{\pm, ijk}(\alpha) \ .
\end{align}
We assume that any additive bias component is well calibrated by lensing image simulations and removed from the measurement pipeline \cite{2022MNRAS.509.3371M}.

Finally, we consider the effect of galaxy IA that describe intrinsic correlations between the shapes of source galaxies and their local tidal fields, i.e.~correlations that are not induced by the gravitational lensing effect. We adopt the {\it nonlinear linear alignment} (NLA) model \cite{2007MNRAS.381.1197H, 2007NJPh....9..444B} for both $\xi_{\pm}$ and $\zeta_{\pm}$. In practice, the incorporation of IA in our theory predictions is equivalent to transforming the lensing kernels as (see App.~\ref{appendix:NLA} for more details)
 \begin{equation}
    \label{eq:lensing_kernel_q_with_NLA}
    q^i(\chi) \longrightarrow q^i(\chi) + f_{\text{IA}}(z(\chi))\frac{n^i_s(\chi)}{\Bar{n}^i_s}\frac{\text{d}z}{\text{d}\chi} \ ,
\end{equation}
with $\Bar{n}^i_s$ the mean source galaxy density in tomographic bin $i$ and \cite{2017arXiv170609359K, 2022PhRvD.106h3509G}
\begin{equation}
    \label{eq:linear_bias_NLA}
    f_{\text{IA}}(z) = -A_{\text{IA},0}\left(\frac{1+z}{1+z_0}\right)^{\alpha_{\text{IA}}}\frac{c_1\rho_{\text{crit}}\Omega_{\text{m},0}}{D(z)} \ ,
\end{equation}
where $A_{\text{IA},0}$ is the IA amplitude, $\alpha_{\text{IA}}$ is a power index and $D(z)$ is the linear growth factor. We adopt $z_0=0.62$, $c_1\rho_{\text{crit}} = 0.0134$ \cite{2001MNRAS.322..419H, 2007NJPh....9..444B}. In our results below we keep the power index fixed to $\alpha_{\text{IA}} = 0$ for simplicity; note that simultaneously varying $A_{\text{IA},0}$ and $\alpha_{\text{IA}}$ in MCMC constraints can lead to posterior projection effects that could artificially bias the marginalized constraints of $A_{\text{IA},0}$ towards zero.

In our modelling of IA, $\xi_{\pm}$ acquires terms $\propto f_{\text{IA}}, f_{\text{IA}}^2$, and $\zeta_{\pm}$ terms $\propto f_{\text{IA}}, f_{\text{IA}}^2, f_{\text{IA}}^3$. These are different from the terms displayed in Ref.~\cite{2021MNRAS.503.2300P}; further notice that our Eq.~(\ref{eq:linear_bias_NLA}) differs from the corresponding Eq.~(27) in Ref.~\cite{2021MNRAS.503.2300P} by a multiplicative factor $1/(1+z)$. We shall return to the impact of different IA treatments when we discuss our numerical results.


\section{Data vector and covariance from simulations}
\label{sec:cov}

In this section we describe the DES Y3-like simulated cosmic shear maps that we use to measure the $\xi_{\pm}$ and $\zeta_{\pm}$ data vectors and to estimate their covariance matrices.

\subsection{Shear maps from $N$-body simulations}
\label{sec:T17}

Our main cosmic shear maps are obtained using the publicly available N-body simulation data developed by Takahashi et al.~\cite{2017ApJ...850...24T} (hereafter referred to as T17). In particular, we make use of the 108 independent full-sky cosmic shear maps for several Dirac-delta source distributions at redshifts between $z = 0.05$ and $z = 5.3$. The cosmology of the simulations is flat $\Lambda$CDM with parameters: $\Omega_m = 0.279$, $\Omega_b = 0.046$, $h = 0.7$, $\sigma_8 = 0.82$, $n_s = 0.97$. 

\begin{figure}
    \centering
    \begin{subfigure}{.5\textwidth}
        \centering
        \includegraphics[width=\linewidth]{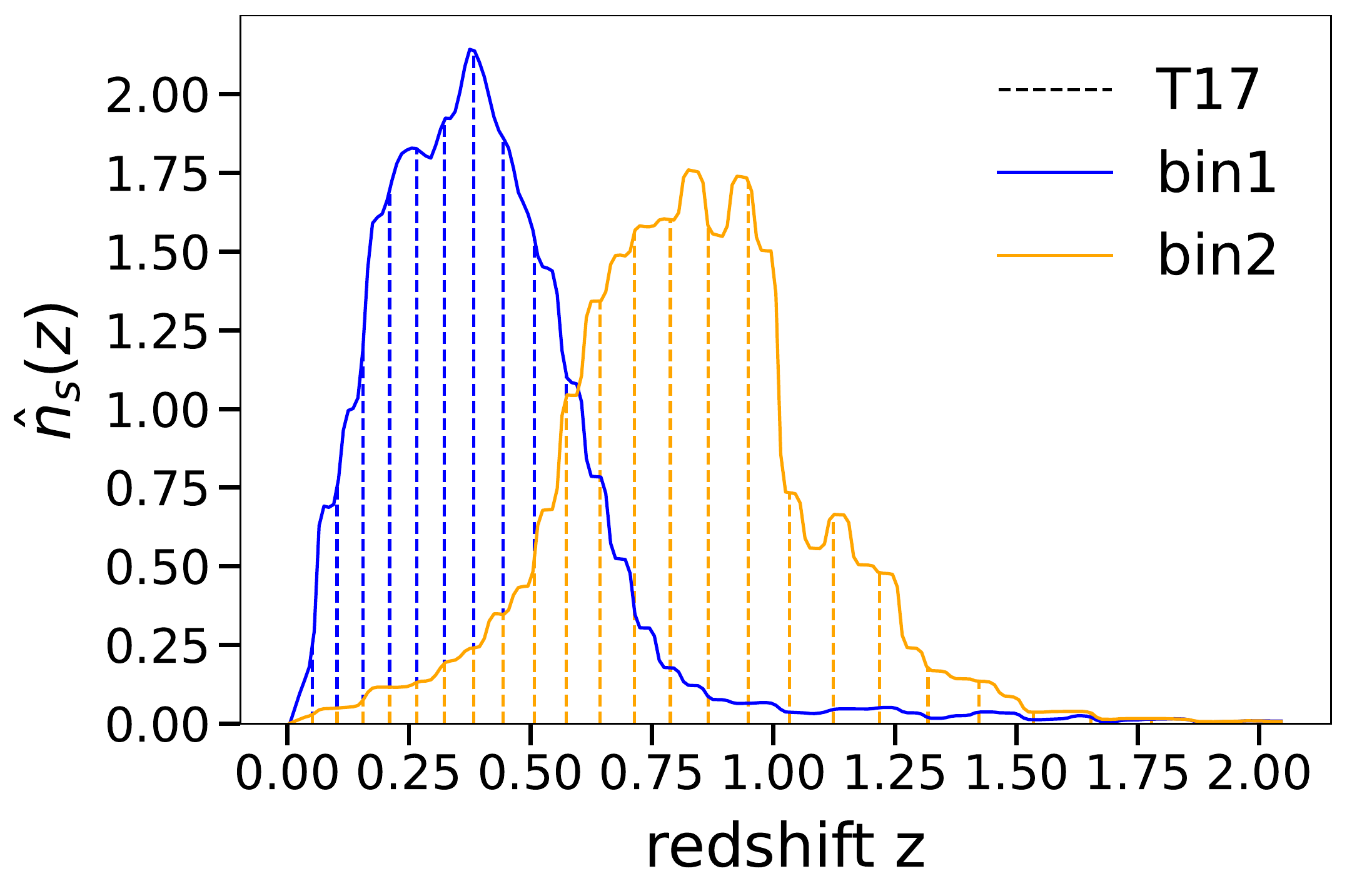}
        \label{fig:merged_nz_and_footprints.a}
    \end{subfigure}%
    \begin{subfigure}{.5\textwidth}
        \centering
        \includegraphics[width=\linewidth]{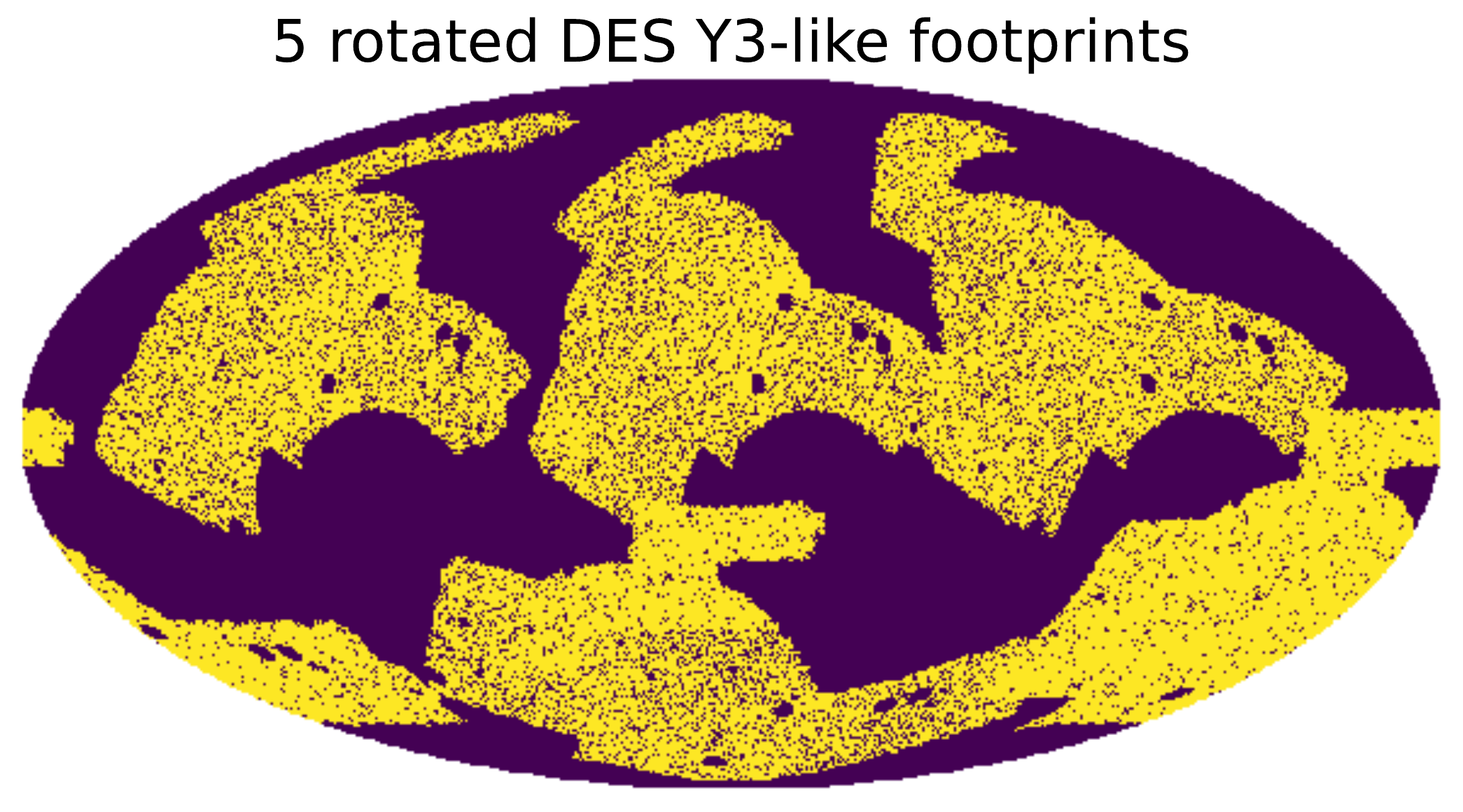}
        \label{fig:merged_nz_and_footprints.b}
    \end{subfigure}
    \caption{\textit{Left panel}: The two galaxy source redshift distributions that we consider in this paper. Each is a combination of two of the four DES Y3 source distributions. The vertical dashed lines mark the source redshifts of the T17 shear maps, which are weighted by the galaxy source distributions to produce our shear maps. \textit{Right panel}: The Mollweide projection map with the placement of 5 DES Y3 survey footprints after the selection with $Q$ filters of $90$ ${\rm arcmin}$ in a full-sky map; dark blue pixels indicate masked/unobserved regions. This allows us to measure 5 DES Y3-like realizations of $\xi_{\pm}$ and $\zeta_{\pm}$ from each full-sky map.}
    \label{fig:merged_nz_and_footprints}
\end{figure}

We consider DES Y3-like galaxy source redshift distributions to construct our cosmic shear maps. For simplicity, rather than considering the four source bins utilized in the DES Y3 analysis, we merge them into two as follows. Let $N_1$ and $N_2$ be the total number of galaxies in the first two DES source distributions $n_{s, \rm DES1}(z)$ and $n_{s, \rm DES2}(z)$, respectively (see Fig.~6 and 11 in Ref.~\cite{2021MNRAS.505.4249M}). Then, our first source redshift bin is obtained as $\hat{n}_s^1 = (N_1 n_{s, \rm DES1} + N_2 n_{s, \rm DES2})/(N_1 + N_2)$; and similarly for our second source redshift, using the third and fourth DES Y3 source distributions. The source redshift distributions that we consider in this paper are shown on the left of Fig.~\ref{fig:merged_nz_and_footprints}. For each of the 108 T17 realizations, we build two full-sky shear maps by summing the T17 shear maps weighted by each of the two source redshift distributions. The vertical lines on the left of Fig.~\ref{fig:merged_nz_and_footprints} mark the source redshift of the T17 maps we use.

We then apply the DES Y3 footprint to each of the full-sky shear maps. In order to maximize the utility of each full-sky map, we place 5 footprints in each with minimal overlap, as illustrated on the right of Fig.~\ref{fig:merged_nz_and_footprints}. For each of our two source bins, this provides us with $108 \times 5 = 540$ DES Y3-like shear maps on which we can measure $\xi_{\pm}$, $\zeta_{\pm}$ and their covariance.

Finally, we add DES Y3 levels of shape noise to our maps as follows. Using the angular positions of the source galaxies in the DES Y3 shape catalogue \cite{2021MNRAS.504.4312G}, we assign to each of our pixels the galaxy ellipticities and measurement weights that are also present in those catalogues. We then randomly rotate the ellipticities of the galaxies assigned to each pixel. The shape noise $\gamma_{\rm noise}$ is the average of these randomly rotated ellipticities weighted by the corresponding measurement weights. This is added to the shear values of the T17 maps $\gamma_{\rm sim}$ to generate the shear measurement in each pixel $\gamma_{\rm pix}$. Concretely, 
\begin{equation}
    \label{eq:shape_noise}
    \gamma_{\rm pix} = \gamma_{\rm noise} + \gamma_{\rm sim} = \frac{\sum_{j = 1}^{N}\omega_j\gamma_{j, \rm DES}\text{exp}(i\phi_j)}{\sum_{j = 1}^{N}\omega_j} + \gamma_{\rm sim} \ ,
\end{equation}
where $N$ is the number of galaxies in a given pixel, $\gamma_{j, \rm DES}$ and $\omega_j$ are the measured ellipticity and weight of the $j$th galaxy and each angle $\phi_j$ is drawn uniformly from $[0, 2\pi]$; note that the average value of $\gamma_{\rm noise}$ across all pixels is zero, but each pixel has in general nonzero values. 

\subsection{Shear maps from lognormal realizations}
\label{sec:flask}

In addition to the T17-based shear maps, we also consider DES Y3-like maps from lognormal lensing realizations generated with the Full-sky Lognormal Astro-fields Simulation Kit \cite{2016MNRAS.459.3693X} (hereafter referred to as FLASK). FLASK takes as input the lensing convergence power spectrum, which we compute theoretically for the T17 cosmology and our two galaxy source redshift distributions. FLASK requires also the value of a logshift parameter, which we obtain by fitting a lognormal probability distribution function (PDF) to the PDF of the T17 maps (see Sec.~4.2 of \cite{2021MNRAS.506.2780H} for more details about the generation of our FLASK shear maps). For each of our two source bins, we generate a total of 300 independent FLASK full-sky cosmic shear maps, on which we place 5 DES Y3-like footprints analogously to the T17 full-sky maps (cf.~right panel of Fig.~\ref{fig:merged_nz_and_footprints}). We add shape noise following the strategy described above for the T17 maps. For each of the two source bins then, we have a total of $5 \times 300 = 1500$ lognormal realizations of a DES Y3-like footprint on which we can measure $\xi_{\pm}$ and $\zeta_{\pm}$.

\subsection{Data vector and covariance measurements}
\label{sec:data_measure}

\begin{figure}
    \centering
    \includegraphics[width=0.8\textwidth]{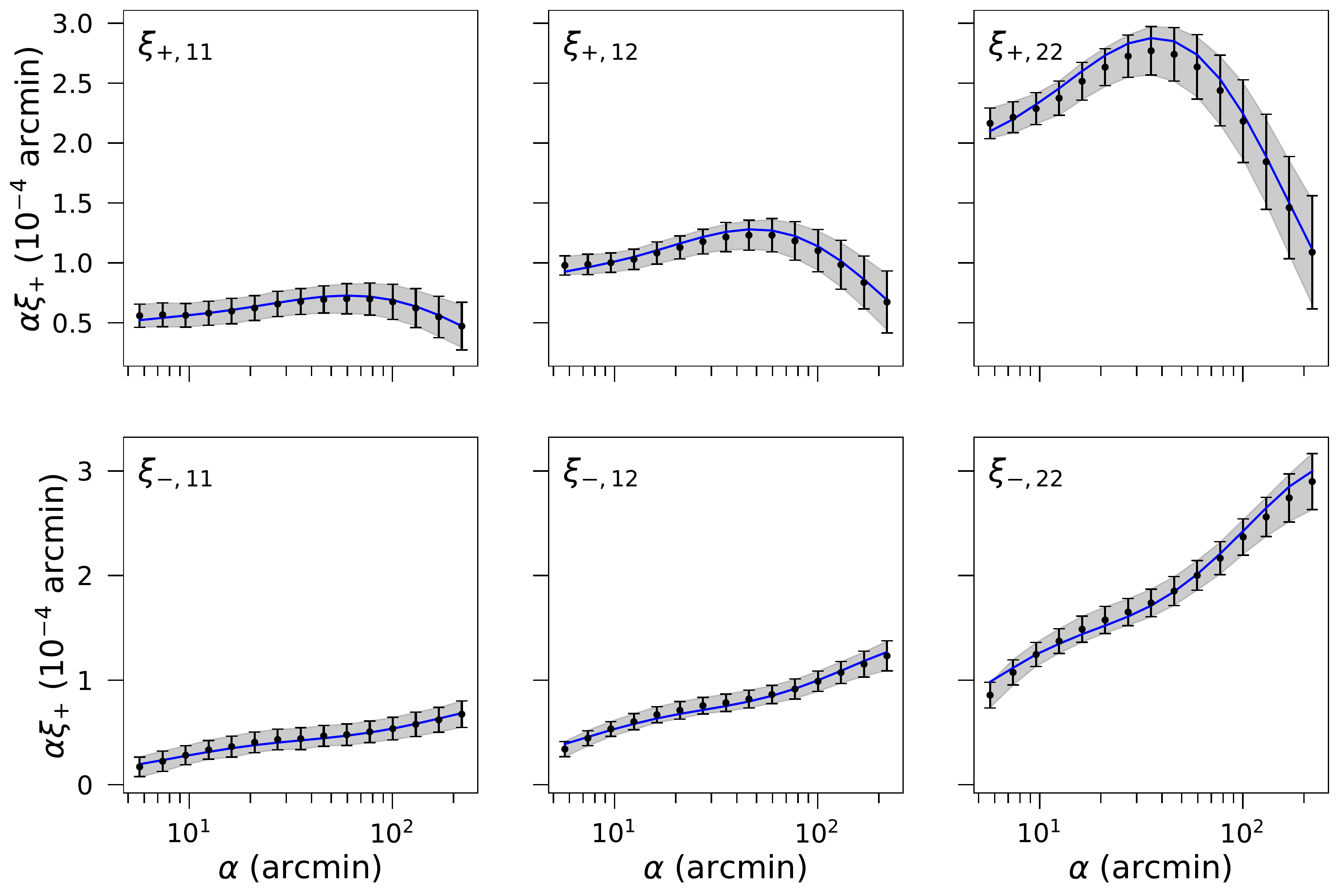}
    \caption{The shear 2PCF $\xi_{\pm}(\alpha)$ measured from our DES Y3-like footprints for two galaxy source redshift bins. The black dots with the error bars show the mean and the standard deviation of the measurements from the $540$ T17 shear maps. For comparison, the grey shaded bands show the standard deviation computed using the $1500$ FLASK shear maps. The blue curves show the theoretical result obtained using Eqs.~\eqref{eq:shear_2pcf_plus} and \eqref{eq:shear_2pcf_minus}.}
    \label{fig:xi_prediction_vs_t17}
\end{figure}

We use the \texttt{Treecorr} code \cite{2004MNRAS.352..338J} to measure $\xi_{\pm, ij}(\alpha)$ on 15 log-spaced angular bins between $5$ and $250\ {\rm arcmin}$; these are scales comparable to those adopted in the DES Y3 analysis \cite{2022PhRvD.105b3515S}. We measure the auto- and cross-correlation of the two source redshift bins, yielding a total of 6 shear 2PCFs. The measurements from the T17 maps are shown by the black dots in Fig.~\ref{fig:xi_prediction_vs_t17}.

\begin{figure}
    \centering
    \includegraphics[width=0.9\textwidth]{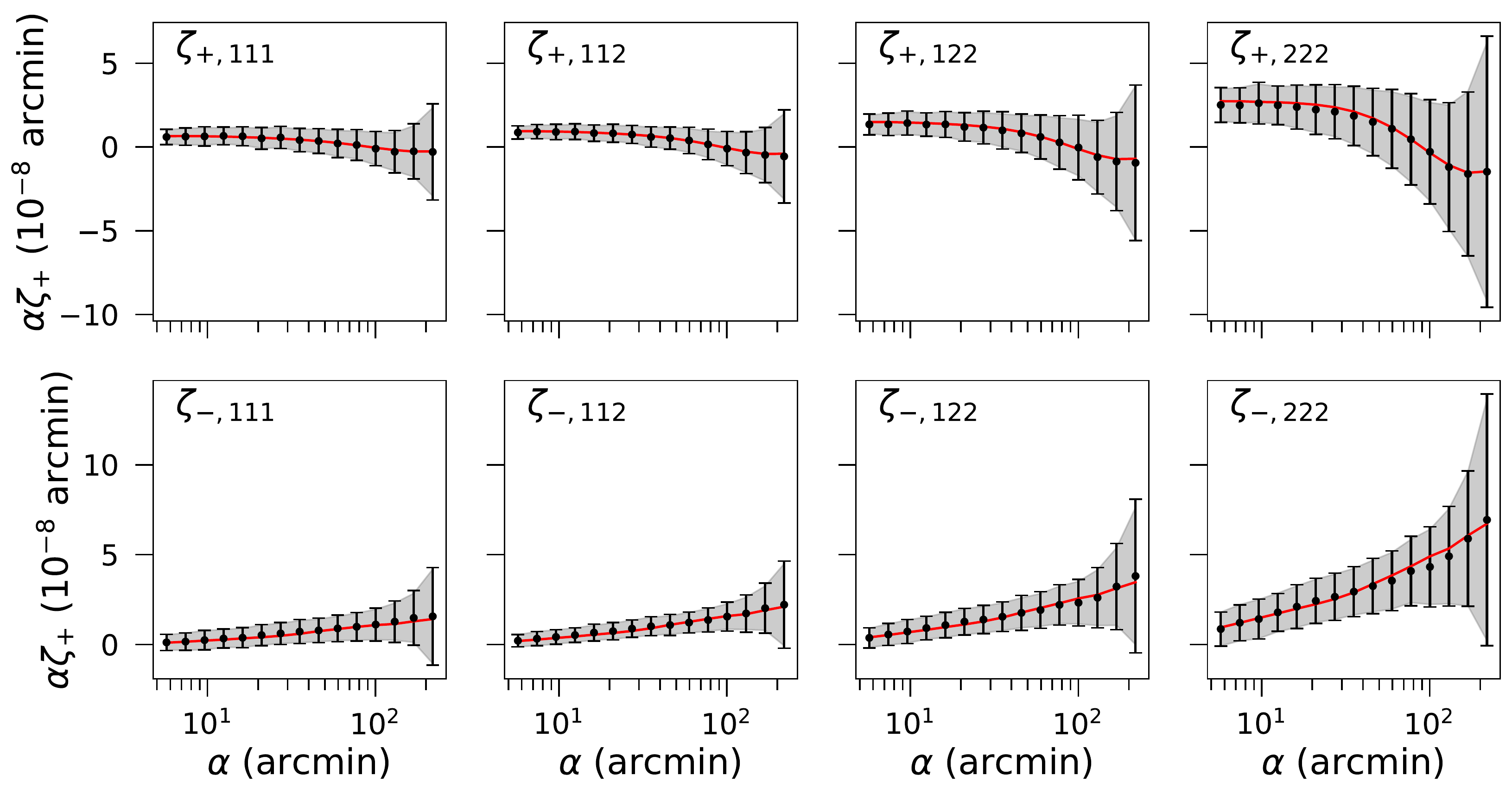}
    \caption{The integrated shear 3PCF $\zeta_{\pm}(\alpha)$ measured from our DES Y3-like footprints for two galaxy source redshift bins. The black dots with the error bars show the mean and the standard deviation of the measurements from the $540$ T17 shear maps for a filter size of $90\ {\rm arcmin}$. The grey shaded bands show the standard deviation computed using the $1500$ FLASK shear maps. The red curves show the theoretical result from Eqs.~\eqref{eq:shear_i3pcf_plus} and \eqref{eq:shear_i3pcf_minus}.}
    \label{fig:zeta_prediction_vs_t17}
\end{figure}

In order to measure $\zeta_{\pm, ijk}(\alpha)$, we use the \texttt{Treecorr} code to compute the position-dependent shear 2PCF and 1-point aperture masses within patches of the footprint; we assume the same size $\theta_{\rm ap}$ and $\theta_{\rm T}$ for the aperture mass and position-dependent 2PCF. The 2PCF in each patch is measured in 15 log-spaced angular bins between $5$ and $2\theta_{\rm T}-10\ {\rm arcmin}$, and the 1-point aperture mass is evaluated using Eq.~\eqref{eq:aperture_mass_Q} with the integral up to $5\theta_{\rm ap}$. The $\zeta_{\pm}$ is obtained by averaging the product of the shear 2PCF and 1-point aperture mass across all patches selected in the footprint. For our two source redshift bins, we have 8 integrated auto- and cross-3PCF $\zeta_{\pm, ijk}(\alpha)$. The measurements from the T17 maps are shown by the black dots in Fig.~\ref{fig:zeta_prediction_vs_t17} for an aperture size of $\theta_{\rm ap} = \theta_{\rm T} = 90\ {\rm arcmin}$.\footnote{As a technical point, in our measurements of $\zeta_{\pm}$ we consider only survey patches where the fraction of unmasked pixels is larger than $80\%$ for the top-hat filter $W$ and larger than $70\%$ for the $Q$ filter up to $5\theta_{\rm ap}$ of aperture radius. Holes and masked pixels inside the footprint contribute to the counting of these fractions, in addition to pixels outside the survey footprint. This ensures our measurements are not affected by too many unmasked pixels in the patches, as confirmed by their excellent agreement with the theory predictions for both $\xi_{\pm}$ and $\zeta_{\pm}$ in Figs.~\ref{fig:xi_prediction_vs_t17} and \ref{fig:zeta_prediction_vs_t17}, respectively.}

We estimate the covariance matrix of our data vectors as
\begin{equation}
    \label{eq:cov_estimation}
    \hat{C} = \frac{1}{N_s - 1}\sum_{i=1}^{N_s}(\hat{d}_i - \hat{d})(\hat{d}_i - \hat{d})^{\rm T} \ ,
\end{equation}
where $N_s$ is the number of footprint realizations ($540$ for T17 and $1500$ for FLASK), $\hat{d}_i$ is the data vector of the $i$-th realization and $\hat{d}$ is the mean data vector across all realizations. When evaluating the inverse covariance matrix, we correct it as 
\begin{equation}
    \label{eq:precision_hartlap}
    \hat{C}^{-1} = \Bigg[\frac{N_{\rm s} - N_{\rm d} - 2}{N_{\rm s} - 1}\Bigg] [1 + A + B(N_{\rm p} + 1)]C^{-1} \  ,
\end{equation}
where
\begin{equation}
    \label{eq:A_factor}
    A = \frac{2}{(N_{\rm s} - N_{\rm d} - 1)(N_{\rm s} - N_{\rm d} - 4)} \ ,
\end{equation}
\begin{equation}
    \label{eq:B_factor}
    B = \frac{N_{\rm s} - N_{\rm d} - 2}{(N_{\rm s} - N_{\rm d} - 1)(N_{\rm s} - N_{\rm d} - 4)} \ ,
\end{equation}
and $N_{\rm d}$ is the size of the data vector ($N_{\rm d} = 90$ for $\xi_{\pm}$, $N_{\rm d} = 120$ for $\zeta_{\pm}$, and $N_{\rm d} = 210$ for their combination), $N_{\rm p}$ is the number of inference parameters and $C^{-1}$ is the directly inverted covariance. The first term in brackets is the bias correction on the inverse covariance from Ref.~\cite{2007A&A...464..399H}, while the second term is a correction factor from Ref.~\cite{2014MNRAS.439.2531P}.

\begin{figure}
    \centering
    \includegraphics[width=\textwidth]{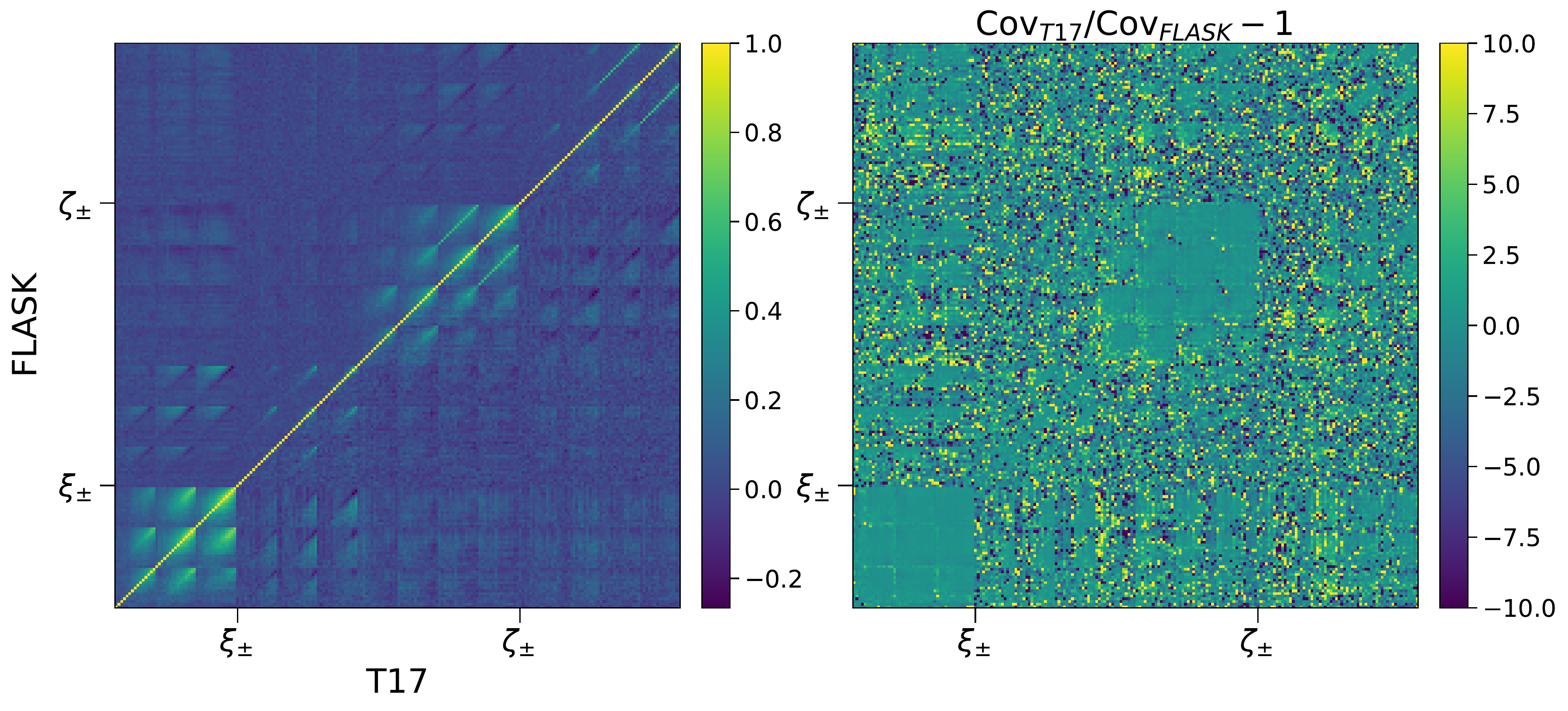}
    \caption{Comparison between the T17 and FLASK covariance matrices. The left panel shows the correlation coefficient from FLASK in the upper triangle and T17 in the lower triangle part of the matrix. The right panel shows the relative difference between the two covariance estimates (the color coding is limited to $\pm10$ to exclude a few extreme values for visibility). The ordering of the matrix entries is according to: $\{\xi_{+,11}$, $\xi_{+,12}$, $\xi_{+,22}$, $\xi_{-,11}$, $\xi_{-,12}$, $\xi_{-,22}$, $\zeta_{+,111}$, $\zeta_{+,112}$, $\zeta_{+,122}$, $\zeta_{+,222}$, $\zeta_{-,111}$, $\zeta_{-,112}$, $\zeta_{-,122}$, $\zeta_{-,222}\}$. The result shown for $\zeta_{\pm}$ is estimated using $90\ {\rm arcmin}$ apertures.}
    \label{fig:corr_cov_T17_flask}
\end{figure}

The FLASK covariance matrix has the advantage of having less numerical noise because of the larger $N_s$, but the disadvantage of corresponding to lognormal realizations of cosmic shear maps, which are not as realistic as the T17 ones from $N$-body simulations. The left panel of Fig.~\ref{fig:corr_cov_T17_flask} compares the correlation matrix $r_{mn} = \hat{C}_{mn}/\sqrt{\hat{C}_{mm}\hat{C}_{nn}}$ from the FLASK (upper triangle) and T17 (lower triangle) maps; the indices ${m, n}$ run over the data vector entries. Reassuringly, the two covariance matrices display broadly the same correlations. There are however some differences that are better seen in the right panel of Fig.~\ref{fig:corr_cov_T17_flask} which shows the relative difference between the two covariances. We investigate the impact of these differences in the parameter constraints when we discuss our results below.

We note that both our covariance matrices do not appropriately account for super-sample covariance (SSC) \cite{2013PhRvD..87l3504T, 2018JCAP...06..015B}, i.e.~the variance induced by the gravitational coupling between observed modes inside the survey and unobserved modes with wavelengths larger than the survey size. The SSC is the dominant off-diagonal contribution in 2-point function analyses \cite{2018JCAP...10..053B}, and it is expected to be a smaller contribution to the squeezed bispectrum configurations that dominate the small-scale $\zeta_{\pm}$ \cite{2019JCAP...03..008B}. Our quoted error bars for $\xi_{\pm}$-only analyses are thus expected to be underestimated, and consequently, our quoted improvements from $\zeta_{\pm}$ are conservative; i.e.~the relative improvement from $\zeta_{\pm}$ is expected to be larger in analyses that appropriately account for SSC. We defer the inclusion of SSC to future work.

\section{Emulators for $\xi_{\pm}$ and $\zeta_{\pm}$}
\label{sec:emulation} 
\begin{table}
    \centering
    \begin{tabular}{|c c|}
        \hline
        & Prior range \\
        \hline
        \hline
        \textbf{Cosmological parameters} (emulated) &  \\
        $\Omega_{\rm m}$ & $U \left[0.16, 0.45\right]$ \\
        ${\rm ln}(10^{10}A_s)$ & $U \left[1.61, 4.20\right]$  \\
        $w_0$ & $U \left[-3.33, -0.33\right]$ \\
        \hline
        \textbf{Baryonic feedback parameter} (emulated) & \\
        $c_{\rm min}$ & $U \left[1.0, 5.5\right]$ \\
        \hline
        \textbf{Systematic parameters} (not emulated) & \\
        $\Delta z_1$ & $\mathcal{N}(0.0, 0.023)$ \\
        $\Delta z_2$ & $\mathcal{N}(0.0, 0.020)$ \\
        $m_1$ & $\mathcal{N}(0.0261, 0.012)$ \\
        $m_2$ & $\mathcal{N}(-0.061, 0.011)$ \\
        $A_{\rm IA,0}$ & $U \left[-5.0, 5.0\right]$ \\
        $\alpha_{\rm IA}$ & $0\ ({\rm fixed})$ \\
        \hline
    \end{tabular}
    \caption{Model parameters considered in this paper. The parameters that enter our NN emulator are the cosmological parameters $\Omega_{\rm m}$, ${\rm ln}(10^{10}A_s)$, $w_0$, and the baryonic feedback parameter $c_{\rm min}$. The photo-$z$, shear calibration and IA systematic parameters do not need to be emulated because the predictions for different values are fast to obtain. In our MCMC analyses we vary these parameters within the listed uniform prior ranges ($U$) or assuming Gaussian priors $\mathcal{N}(\mu, \sigma)$ with mean $\mu$ and standard deviation $\sigma$. The listed priors for the systematic parameters are inspired by those assumed in the DES Y3 analyses \cite{2022PhRvD.106h3509G, 2022PhRvD.105b3515S}.}
    \label{tab:param_prior}
\end{table}

The evaluation of the integrated lensing bispectrum $\mathcal{B}^{\rm 2D}_{\pm, ijk}(\boldsymbol{\ell})$ is the key computational bottleneck when evaluating $\zeta_{\pm}$ using Eqs.~(\ref{eq:shear_i3pcf_plus}) and (\ref{eq:shear_i3pcf_minus}), and thus the quantity that we wish to emulate. However, rather than emulating $\mathcal{B}^{\rm 2D}_{\pm, ijk}(\boldsymbol{\ell})$ directly, we emulate only the part of the integrand in Eq.~(\ref{eq:integrated_bispectrum_2d}) given by
\begin{equation}
    \label{eq:emulated_quantity}
    \begin{split}
    \int \frac{\text{d}^2\boldsymbol{\ell}_1}{(2\pi)^2} \int\frac{\text{d}^2\boldsymbol{\ell}_2}{(2\pi)^2} B^{\rm 3D}_{\delta}\left(\frac{\boldsymbol{\ell}_1}{\chi}, \frac{\boldsymbol{\ell}_2}{\chi}, \frac{-\boldsymbol{\ell}_1-\boldsymbol{\ell}_2}{\chi}, \chi\right) \times e^{2i(\phi_2 \mp \phi_{-1-2})}U(\boldsymbol{\ell}_1)W(\boldsymbol{\ell}_2+\boldsymbol{\ell})W(-\boldsymbol{\ell}_1-\boldsymbol{\ell}_2-\boldsymbol{\ell}).
    \end{split}
\end{equation}
This leaves out the part involving the line-of-sight integration in Eq.~(\ref{eq:integrated_bispectrum_2d}), but has the advantage of allowing for more flexibility to adjust the source redshift distributions, including bypassing the need to emulate any of the systematic parameters mentioned in Sec.~\ref{sec:systematics}. The training of the emulator still needs to be redone for different sizes of the $U$ and $W$ filters. The direct evaluation of $\xi_{\pm}$ in an MCMC exploration of the parameter space would not impose a serious computational burden, but we emulate its calculation anyway for extra speed. In this case we emulate simply the three-dimensional matter power spectrum $P^{3D}_{\delta}$ in Eq.~(\ref{eq:kappa_power_spectrum}).

We build our emulator by training a neural network (NN) on a Latin hypercube with $10^{5}$ training nodes. The emulated parameters comprise the cosmological parameters $\{\Omega_{\rm m}, A_s, w_0\}$, the baryonic feedback parameter $c_{\rm min}$, as well as the redshift $z$ which we need to emulate to perform the line-of-sight integrations in Eqs.~(\ref{eq:integrated_bispectrum_2d}) and (\ref{eq:kappa_power_spectrum}). The ranges of the cosmological and baryonic parameters are listed in Tab.~\ref{tab:param_prior} (note we rescale $A_s$ to ${\rm ln}(10^{10}A_s)$), and for redshift we consider $z \in \left[0, 2.1\right]$. The NN architecture is that of the \texttt{Cosmopower} code \cite{2022MNRAS.511.1771S}\footnote{\url{https://alessiospuriomancini.github.io/cosmopower/}}, which was originally developed to emulate 2-point statistics, but which can be straightforwardly applied to emulate Eq.~(\ref{eq:emulated_quantity}). The input layers  of the NN are the cosmological, baryonic and redshift parameters. For $\zeta_{\pm}$, the output of the NN is the quantity in Eq.~(\ref{eq:emulated_quantity}) in $100$ log-spaced $\ell$ bins between $\ell=2$ and $\ell=15000$. In the training set, the supervised learning labels are the same quantity obtained by directly evaluating Eq.~(\ref{eq:emulated_quantity}) using Monte-Carlo integration. For $\xi_{\pm}$ the output is the three-dimensional matter power spectrum in $100$ log-spaced $\ell$ bins as in the right-hand side of Eq.~(\ref{eq:kappa_power_spectrum}) between $\ell = 2$ and $\ell = 15000$.

We test the emulators using another Latin hypercube with $10^{3}$ test nodes with the same prior ranges of the training set. We quantify the performance of the emulator with the expression
\begin{equation}
    \label{eq:test_chi2_fractional_difference}
    \epsilon \equiv \left|\frac{\chi^2_{{\rm emu},i}}{\chi^2_{{\rm test},i}} - 1\right|\ ,
\end{equation}
where $\chi^2_{{\rm emu}, i}$ is the $\chi^2$ value associated with the $i$th test node, defined w.r.t.~the data vector $\hat{d}_{\rm T17}$ generated by the theory model at T17 cosmological parameters. Concretely, $\chi^2_{{\rm emu}, i} = \left(\hat{d}_{{\rm emu}, i} - \hat{d}_{\rm T17}\right)^t\hat{C}^{-1}\left(\hat{d}_{{\rm emu}, i} - \hat{d}_{\rm T17}\right)$, with $\hat{d}_{{\rm emu}, i}$ the emulator prediction and $\hat{C}^{-1}$ the T17 inverse covariance matrix. The quantity $\chi^2_{{\rm test}, i}$ is defined analogously, but replacing the emulator result at each test node with the test label prediction. The $\epsilon$ metric describes how similar the emulator would behave to the theory model in likelihood analyses. The smaller the value of $\epsilon$, the better the accuracy of the emulator.

\begin{figure}
	\centering
	\includegraphics[width=\textwidth]{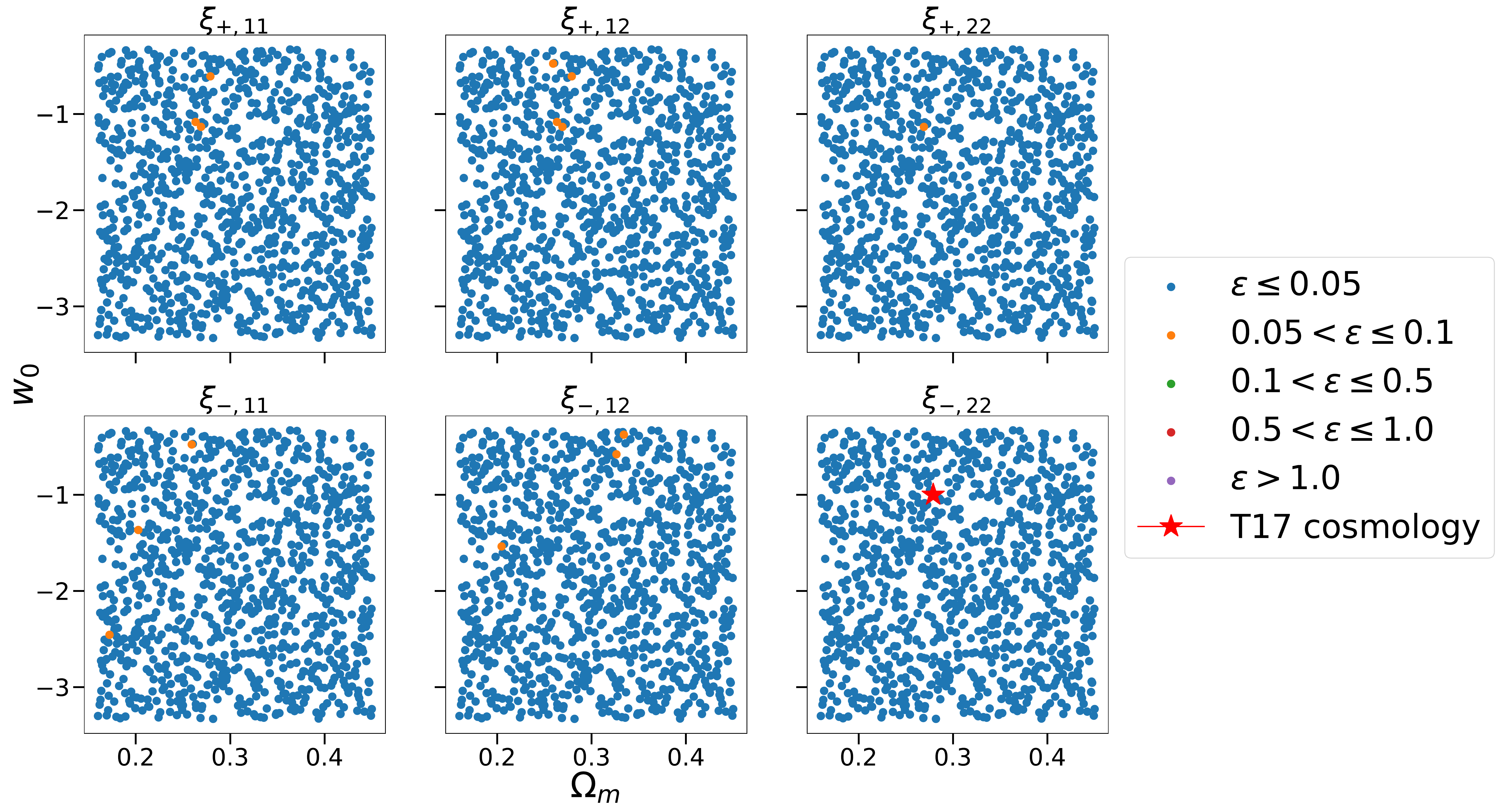}
	\caption{Performance of the $\xi_{\pm}$ emulator on $10^3$ test nodes projected on the $\Omega_{\rm m}$ - $w_0$ plane. The colors show the absolute value of the $\chi^{2}$ relative difference $\epsilon$ defined in Eq.~(\ref{eq:test_chi2_fractional_difference}); if $\epsilon < 0.05$, this means the emulator describes the $\chi^2$ w.r.t.~the T17 cosmology (marked by the red star) to better than $5\%$.} 
    \label{fig:chi2_fractional_difference_xi}
\end{figure}

\begin{figure}
    \centering
    \includegraphics[width=\textwidth]{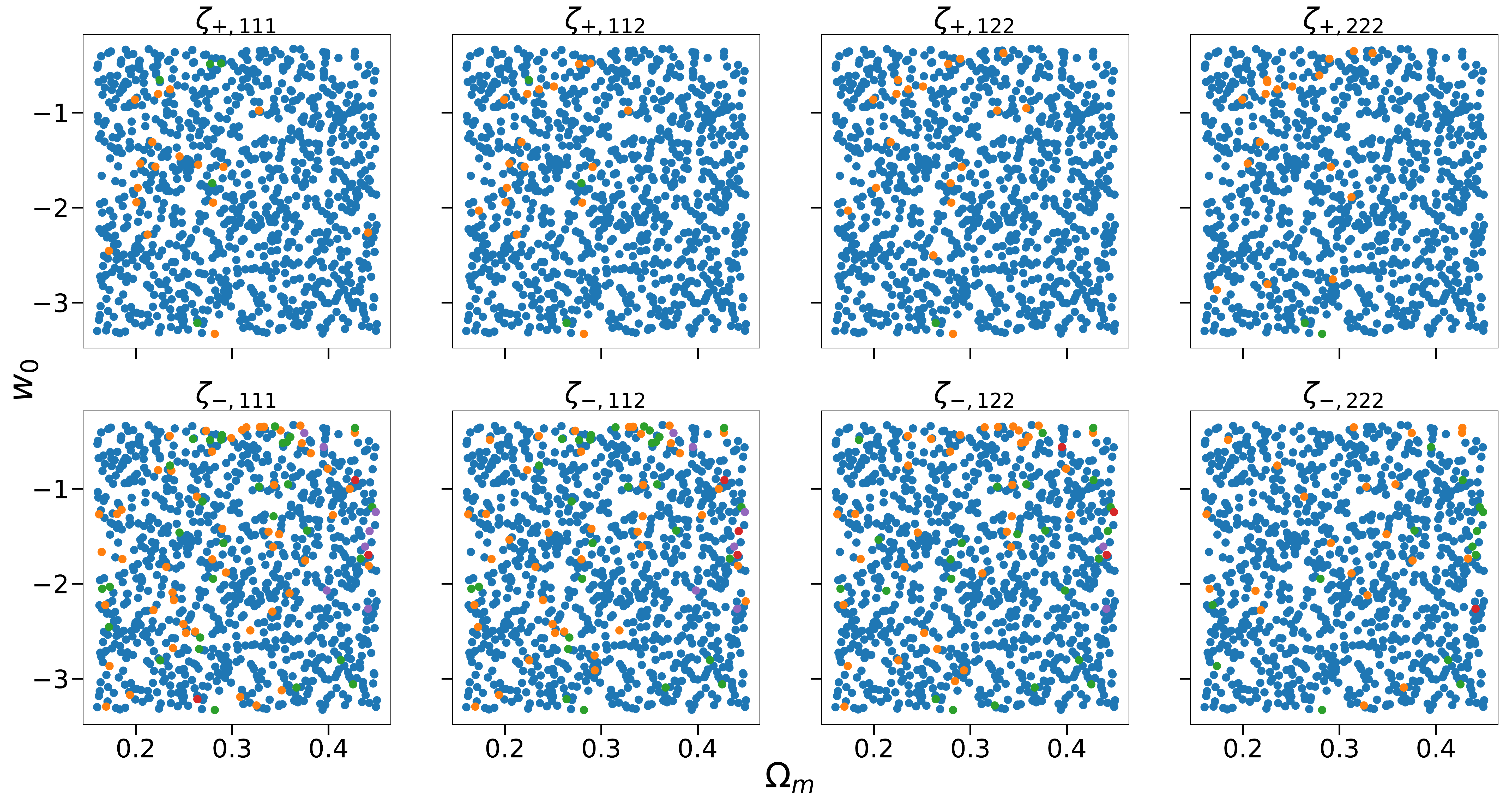}
    \caption{Same as Fig.~\ref{fig:chi2_fractional_difference_xi}, but for the performance of the $\zeta_{\pm}$ emulator, instead of $\xi_{\pm}$. The result is for $90\ {\rm arcmin}$ apertures. The color coding is the same as in Fig.~\ref{fig:chi2_fractional_difference_xi}.}\label{fig:chi2_fractional_difference_zeta}
\end{figure}

Figures \ref{fig:chi2_fractional_difference_xi} and \ref{fig:chi2_fractional_difference_zeta} show the outcome of this test for $\xi_{\pm}$ and $\zeta_{\pm}$, respectively. We show $\epsilon$ projected only on the $\Omega_{\rm m}$ - $w_0$ plane, but the takeaways are common to other projections. For $\xi_{\pm}$, effectively all of the test nodes have $\chi^2$ relative differences $\epsilon < 0.05$. The performance gets reduced slightly for $\zeta_{\pm}$ with $92\%$ ($95\%$) of the test nodes having $\epsilon < 0.05$ ($\epsilon < 0.1$); the result in Fig.~\ref{fig:chi2_fractional_difference_zeta} is for apertures with $90\ {\rm arcmin}$, but we have checked the performance is equivalent for other apertures as well. If the true $\chi^2$ value of some point in parameter space is $\chi^{2}_{\rm test} = 1$, then $\epsilon < 0.1$ implies $\chi^2_{\rm emu} \in \left[0.9, 1.1\right]$. Effectively all of the test nodes for both $\xi_{\pm}$ and $\zeta_{\pm}$ satisfy this satisfactory criterion.

\section{Results: simulated likelihood analyses with MCMC}
\label{sec:results}

In this section we present our main numerical results from simulated likelihood analyses with MCMC. Unless otherwise specified, we consider the parameter priors listed in Tab.~\ref{tab:param_prior}, and sample the parameter space assuming a Gaussian likelihood function,
\begin{equation}\label{eq:Glike}
\mathcal{L}(\boldsymbol{\theta}) \propto {\rm exp}\left[-\frac{1}{2}\left(\mu(\boldsymbol{\theta}) - \hat{d}\right)^t C^{-1} \left(\mu(\boldsymbol{\theta}) - \hat{d}\right)\right]\ ,
\end{equation}
where $\hat{d}$ is the assumed data vector, $C$ the covariance matrix and $\mu(\boldsymbol{\theta})$ the theory prediction for model parameters $\boldsymbol{\theta}$. We utilize the sampler code \texttt{affine}\footnote{\url{https://github.com/justinalsing/affine}} based on \texttt{tensorflow}. With the available NVIDIA A100 GPU (Graphics Processing Unit) hardware, emulator and sampler, we are able to sample an order of $10^6$ points in an hour's timescale.

Next, we validate our model using the T17 $\xi_{\pm}$ and $\zeta_{\pm}$ data vectors in Sec.~\ref{sec:novel_bayesian}, investigate the impact of the aperture size in $\zeta_{\pm}$ constraints in Sec.~\ref{sec:filter_optimisation}, discuss the impact of the systematic parameters in Sec.~\ref{sec:self_calibra_nla}, and check the impact from using the T17 or FLASK covariance matrices in Sec.~\ref{sec:cov_impact_param_inference}. All of the marginalized two-dimensional constraints shown throughout display contours with the $1\sigma$ and $2\sigma$ confidence regions.
\subsection{Validation on the T17 cosmic shear maps}
\label{sec:novel_bayesian}

\begin{figure}
    \centering
    \includegraphics[scale=0.35]{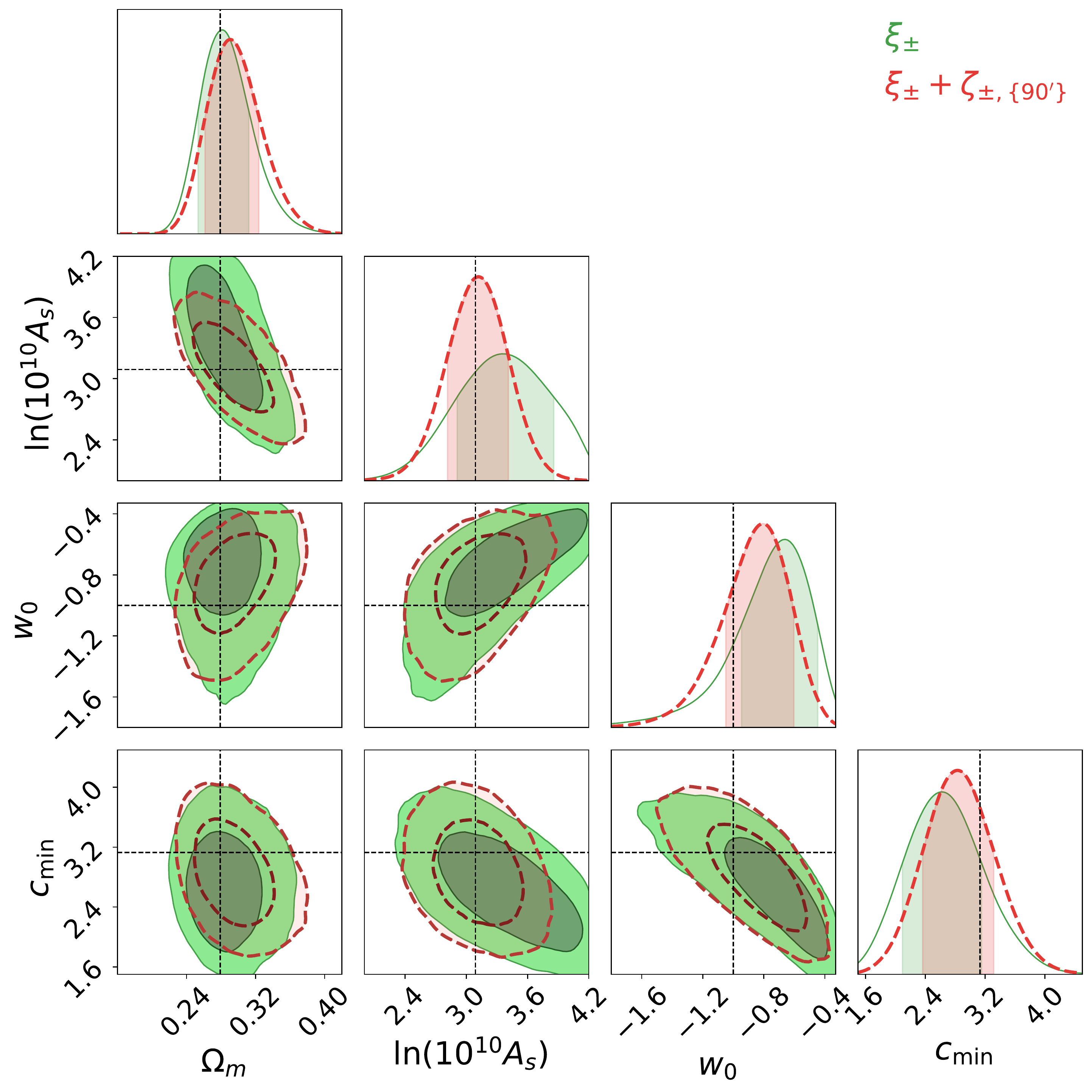}
     \caption{Parameter constraints obtained with the T17 data vector and FLASK covariance matrix. The constraints in green and red are for the $\xi_{\pm}$-only and $\xi_{\pm} + \zeta_{\pm}$ data vectors, respectively. The black dashed lines mark the T17 parameters.}
    \label{fig:model_validation}
\end{figure}

Figure \ref{fig:model_validation} shows the constraints on the cosmological and baryonic feedback parameters for the data vector from the T17 shear maps (cf.~black points in Figs.~\ref{fig:xi_prediction_vs_t17} and \ref{fig:zeta_prediction_vs_t17}) and the FLASK covariance matrix. The result is for $\zeta_{\pm}$ measured using $90$ ${\rm arcmin}$ apertures. We keep the systematic parameters fixed to zero in these constraints, which is the case for our T17 maps. In addition to the correction factors in Eq.~\eqref{eq:precision_hartlap}, in this section we consider also the factor $\left[1 + B(N_{\rm d} - N_{\rm p})\right]^{-1}$ from Ref.~\cite{2013PhRvD..88f3537D} due to statistical noise in our covariance matrix estimate.

The key takeaway from Fig.~\ref{fig:model_validation} is that our theory model and emulator recover unbiased constraints: the T17 parameters (dashed black lines) are contained well within the $1\sigma$ confidence levels for both the $\xi_{\pm}$-only (green) and $\xi_{\pm}+\zeta_{\pm}$ constraints (red). The ability of our theory model to recover unbiased cosmological constraints could have already been anticipated from the good agreement between theory and simulations in Figs.~\ref{fig:xi_prediction_vs_t17} and \ref{fig:zeta_prediction_vs_t17}. 

As a test, we have repeated the analysis in Fig.~\ref{fig:model_validation} but adopting the t-distribution likelihood function from Ref.~\cite{2022MNRAS.510.3207P}, instead of a Gaussian likelihood. The result (not shown) is practically indistinguishable from that in Fig.~\ref{fig:model_validation} for both $\xi_{\pm}$ and $\xi_{\pm}+\zeta_{\pm, \{90^{\prime}\}}$, suggesting the exact choice of the likelihood function does not critically affect our results.

\subsection{The impact of the aperture size}
\label{sec:filter_optimisation}

\begin{table}
    \centering
    \begin{tabular}{|c|c|c|c|c|}
        \hline
        Aperture sizes (${\rm arcmin}$) & $\Omega_{\rm m}$ & ${\rm ln}\left(10^{10}A_s\right)$ & $w_0$ & $c_{\rm min}$ \\
        \hline
        \hline 
        $50$ & 1.2\% & 9.0\% & 18.1\% & 4.8\% \\
        $70$ & 1.2\% & 16.9\% & 31.9\% & 11.6\% \\
        $90$ & \textbf{3.7}\% & \textbf{20.2}\% & \textbf{38.4}\% & \textbf{15.1}\% \\
        $110$ & 1.2\% & 19.1\% & 34.1\% & 11.0\% \\
        $130$ & 1.2\% & 16.9\% & 32.6\% & 12.3\% \\
        \hline 
	\hline 
        $\{50, 70, 90\}$ & 2.5\% & 24.7\% & 39.1\% & 15.8\% \\
        $\{50, 90, 130\}$ & 3.7\% & 23.6\% & 41.3\% & 16.4\% \\
        $\{70, 90, 110\}$ & 6.2\% & 25.8\% & 39.1\% & 15.1\% \\
        $\{90, 110, 130\}$ & 8.6\% & 25.9\% & 42.8\% & 15.8\% \\
        $\{50, 70, 90, 110, 130\}$ & \textbf{12.4}\% & \textbf{28.1}\% & \textbf{44.9}\% & \textbf{19.9}\% \\
        \hline
    \end{tabular}
    \caption{Relative improvement of combined $\xi_{\pm}+\zeta_{\pm}$ constraints relative to $\xi_{\pm}$-only for different values and combinations of the aperture sizes. The best single- and combined-filter cases are highlighted in bold. The result is for a noiseless data vector from the theory model, the FLASK covariance, and marginalizing over the systematic parameters.}
    \label{tab:single_and multiple_filter_param_improve}
\end{table}

\begin{figure}
    \centering
    \includegraphics[scale=0.35]{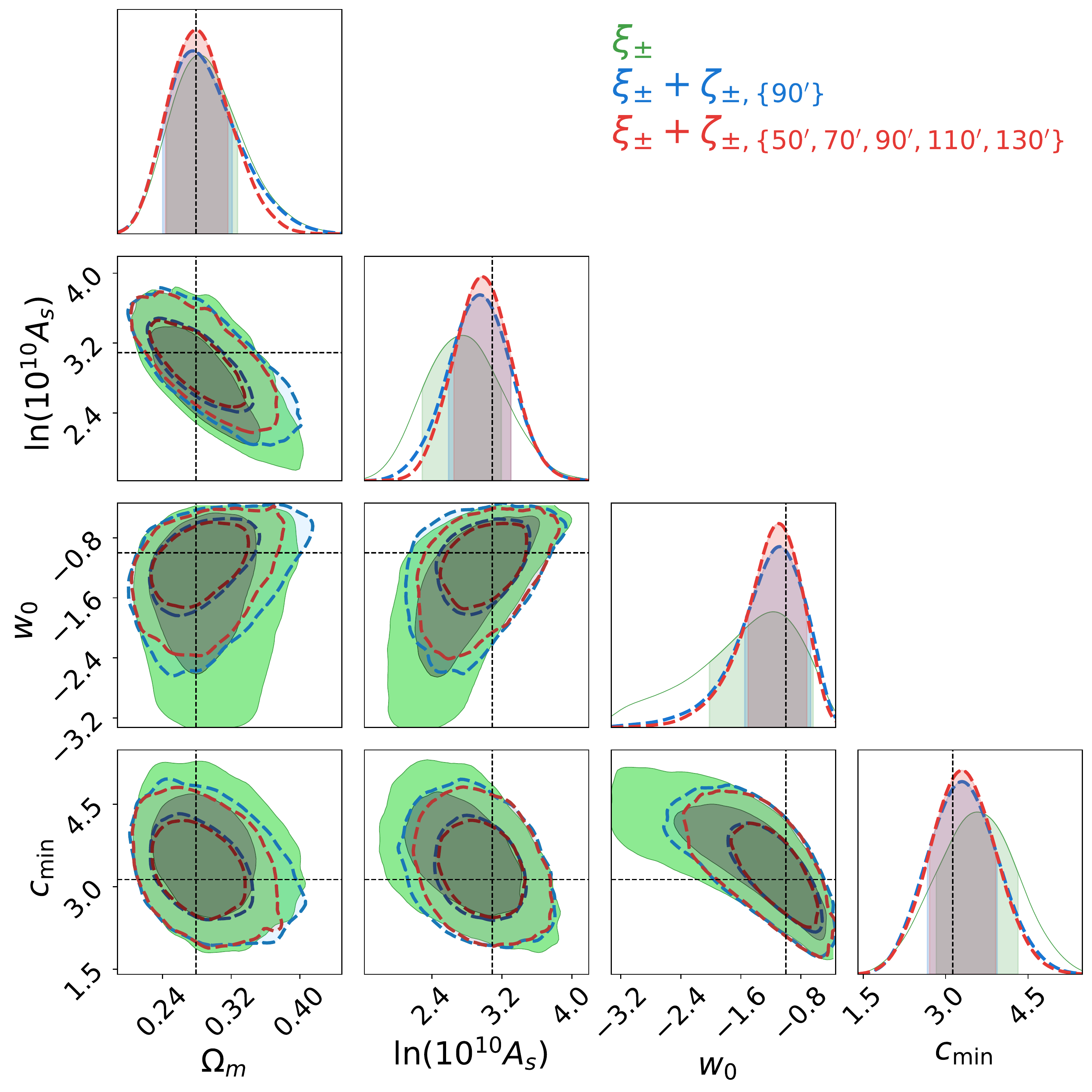}
    \caption{Impact of the aperture size in $\xi_{\pm}+\zeta_{\pm}$ constraints. The contours in green are for $\xi_{\pm}$-only constraints. The contours in blue are for $\xi_{\pm}+\zeta_{\pm}$ constraints using a single aperture with size $90\ {\rm arcmin}$, and in red for the combination of five filter sizes $\{50, 70, 90, 110, 130\}\ {\rm arcmin}$. The result is for a noiseless data vector from the theory model with the T17 parameters (dashed lines), the FLASK covariance, and marginalizing over the systematic parameters.}
    \label{fig:single_and_multiple_filter_optimise}
\end{figure}

When measuring $\zeta_{\pm}$ one of the decisions concerns the choice of the apertures on which to measure the 1-point shear aperture mass and local $\xi_{\pm}$. To investigate the impact of this, we perform likelihood analyses with a noiseless data vector generated with the theory model using the T17 parameters. In these tests, we use the FLASK covariance, and vary also the systematic parameters with the priors listed in Tab.~\ref{tab:param_prior}. The main result is shown in Tab.~\ref{tab:single_and multiple_filter_param_improve}, which lists the relative improvement of the combined $\xi_{\pm}+\zeta_{\pm}$ constraints relative to $\xi_{\pm}$-only, for different aperture sizes and combinations. Figure \ref{fig:single_and_multiple_filter_optimise} shows the actual parameter constraints for two aperture choices: a single aperture with $90\ {\rm arcmin}$ (blue) and the combination of five apertures with sizes $\{50, 70, 90, 110, 130\}\ {\rm arcmin}$ (red).

Regarding the single aperture cases, Tab.~\ref{tab:single_and multiple_filter_param_improve} shows that the constraints improve first from $50$ to $90\ {\rm arcmin}$, but then degrade from $90$ to $130\ {\rm arcmin}$. This follows from the combination of the following effects. Smaller apertures have the advantage of providing $\zeta_{\pm}$ with higher signal-to-noise ratio since there are more apertures over which the average of Eq.~(\ref{eq:i3pcf_ensemble_average}) can be taken. They have, however, the disadvantage that the local $\xi_{\pm}$ is measured over a more reduced range of angular scales inside each patch. Conversely, bigger apertures allow to probe the local $\xi_{\pm}$ on larger scales, but at the price of less signal-to-noise as one averages over a smaller number of patches on the sky.\footnote{In particular, in the limit of very large apertures, the $\xi_{\pm}$ measured in the patches become almost perfectly correlated with the $\xi_{\pm}$ of the whole survey, effectively contributing with no independent information.} In general, different aperture sizes are sensitive to different configurations of the small-scale squeezed-limit bispectrum \cite{2022MNRAS.515.4639H}, which can contain varying cosmological information and impact the final parameter constraints. 

For the aperture sizes shown, the balance between these effects is optimal for apertures with $90\ {\rm arcmin}$, which gives the best constraints. Concretely, the addition of $\zeta_{\pm}$ to the constraints leads to improvements of $4\%$ for $\Omega_{\rm m}$, $20\%$ for ${\rm ln}\left(10^{10}A_s\right)$, $38\%$ for $w_0$ and $15\%$ for $c_{\rm min}$. These figures are in line with the previous findings of Refs.~\cite{2021MNRAS.506.2780H, 2022MNRAS.515.4639H} based on idealized Fisher-matrix forecasts, but extended here to more realistic MCMC-based analyses.

The $\zeta_{\pm}$ measured over slightly different aperture sizes are expected to be substantially correlated due to the large overlap of the regions where the local $\xi_{\pm}$ is measured. However, the lower part of Tab.~\ref{tab:single_and multiple_filter_param_improve} shows that there is still enough independent information to improve the constraints further by combining different apertures. For the cases shown, the best constraints are obtained when combining all apertures $\{50, 70, 90, 110, 130\}\ {\rm arcmin}$: the improvements become $12\%$ for $\Omega_{\rm m}$, $28\%$ for ${\rm ln}\left(10^{10}A_s\right)$, $45\%$ for $w_0$ and $20\%$ for $c_{\rm min}$. These improvements need however to be contrasted with the complications that they add to the analyses. For example, this comes with the price of a much larger data vector, which puts pressure on the numerical requirements for reliable covariance estimates from simulations. In this paper, this pressure was still manageable for a DES Y3-like survey with two tomographic bins, but future survey analysis settings will have larger areas and more source redshift bins as well. The decision of how many filters to combine should thus be made case by case.

\subsection{The impact of systematics and their modelling}
\label{sec:self_calibra_nla}

\begin{figure}
    \centering
    \begin{subfigure}{.5\textwidth}
        \centering
 	\includegraphics[width=\linewidth]{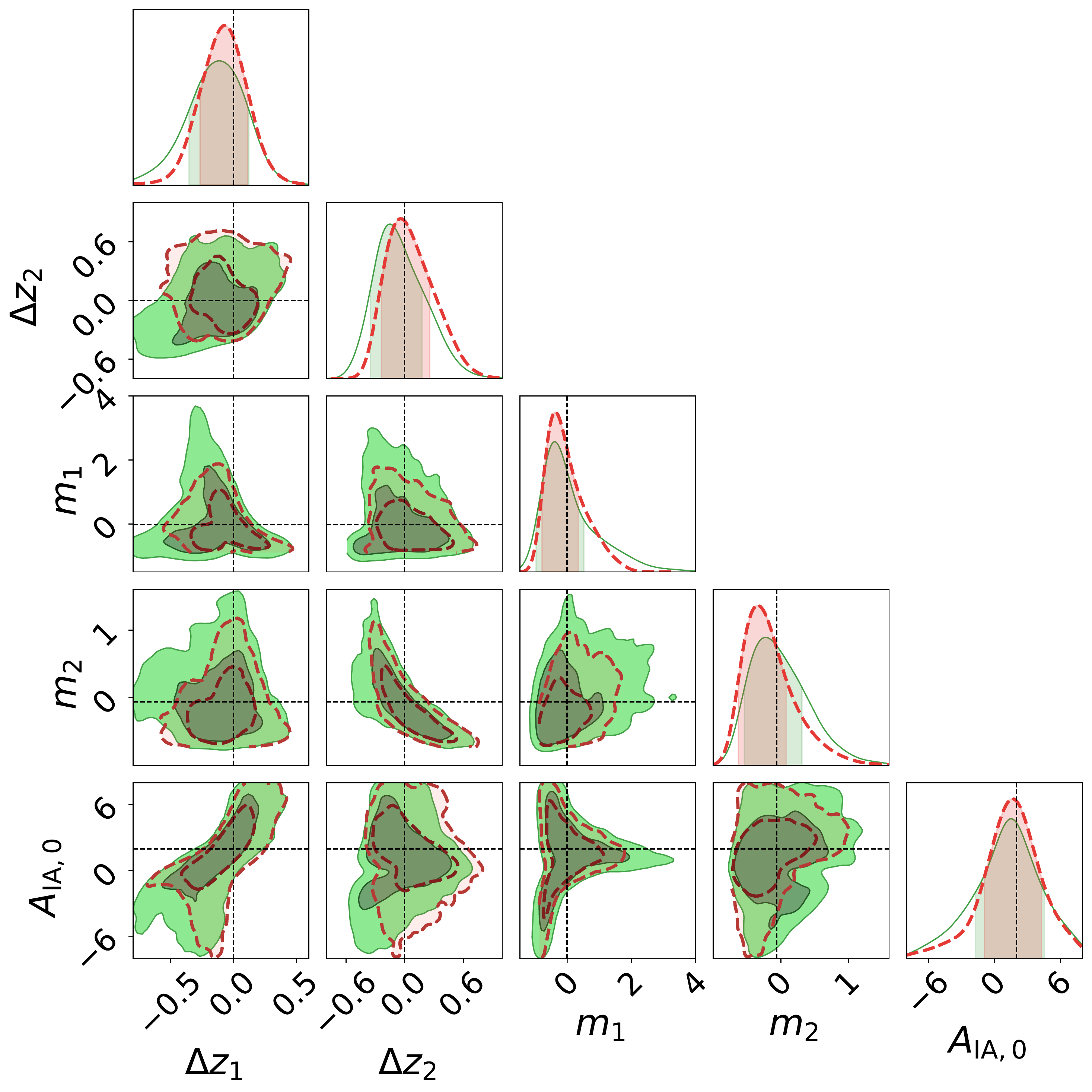}
  	\label{fig:self_calibration.a}
    \end{subfigure}%
    \begin{subfigure}{.5\textwidth}
        \centering
  	\includegraphics[width=\linewidth]{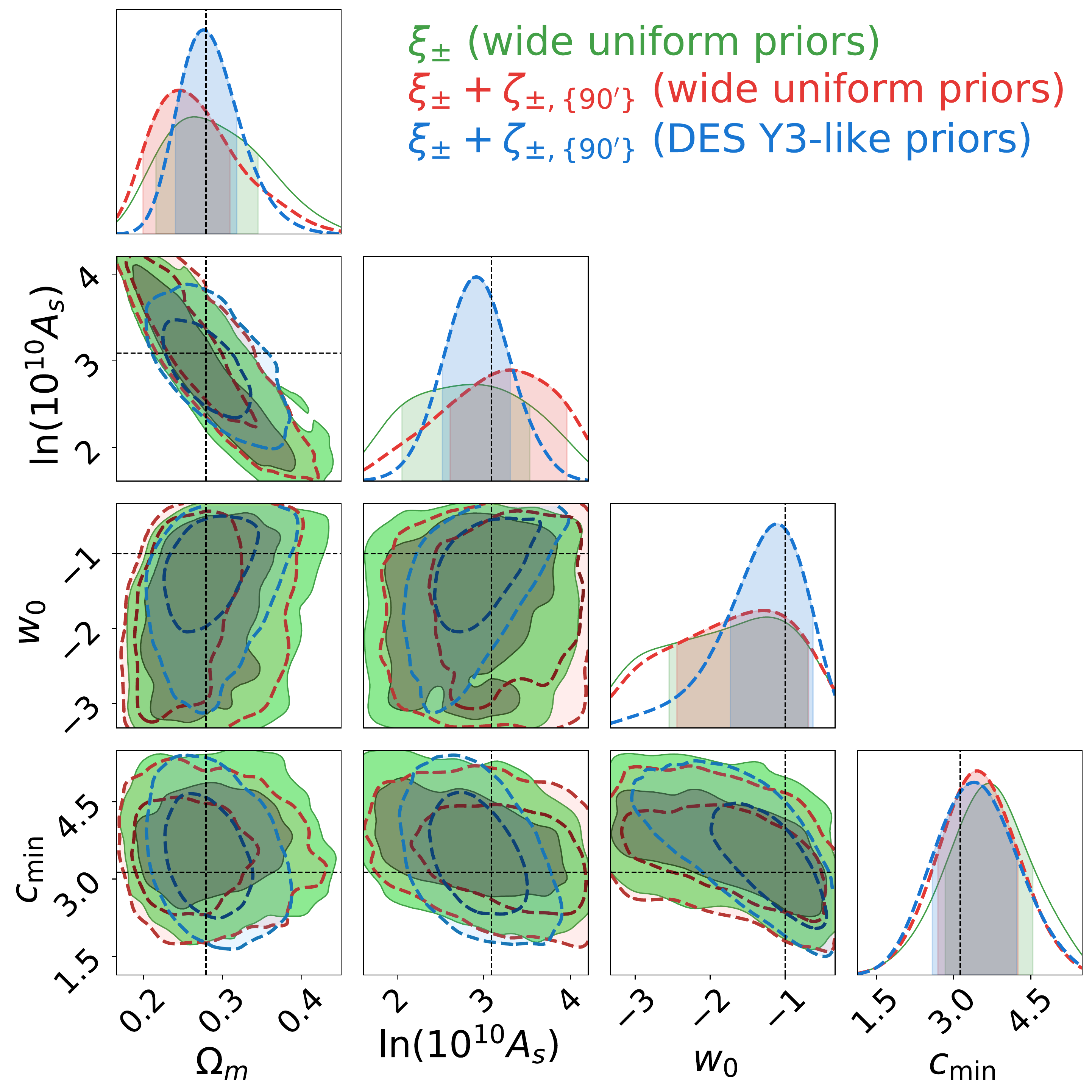}
  	\label{fig:self_calibration.b}
    \end{subfigure}
    \caption{Parameter constraints for different priors on the systematic parameters. The green and red contours are for $\xi_{\pm}$ and $\xi_{\pm} + \zeta_{\pm}$ assuming wide uniform priors on the systematic parameters. The blue contours are for $\xi_{\pm} + \zeta_{\pm}$ with DES Y3-like Gaussian priors on the systematic parameters. The left panel shows the systematic parameter constraints, and the right panel shows the constraints on the cosmological and baryonic feedback parameters. The left panels do not show the contours with DES Y3-like priors as they are too small to be clearly seen. The result is for a noiseless data vector drawn from the theory model with the T17 parameters (except we set $A_{\rm IA, 0} = 2$; cf.~dashed lines), and using the FLASK covariance.}
    \label{fig:self_calibration}
\end{figure}

\begin{figure}
    \centering
    \includegraphics[scale=0.45]{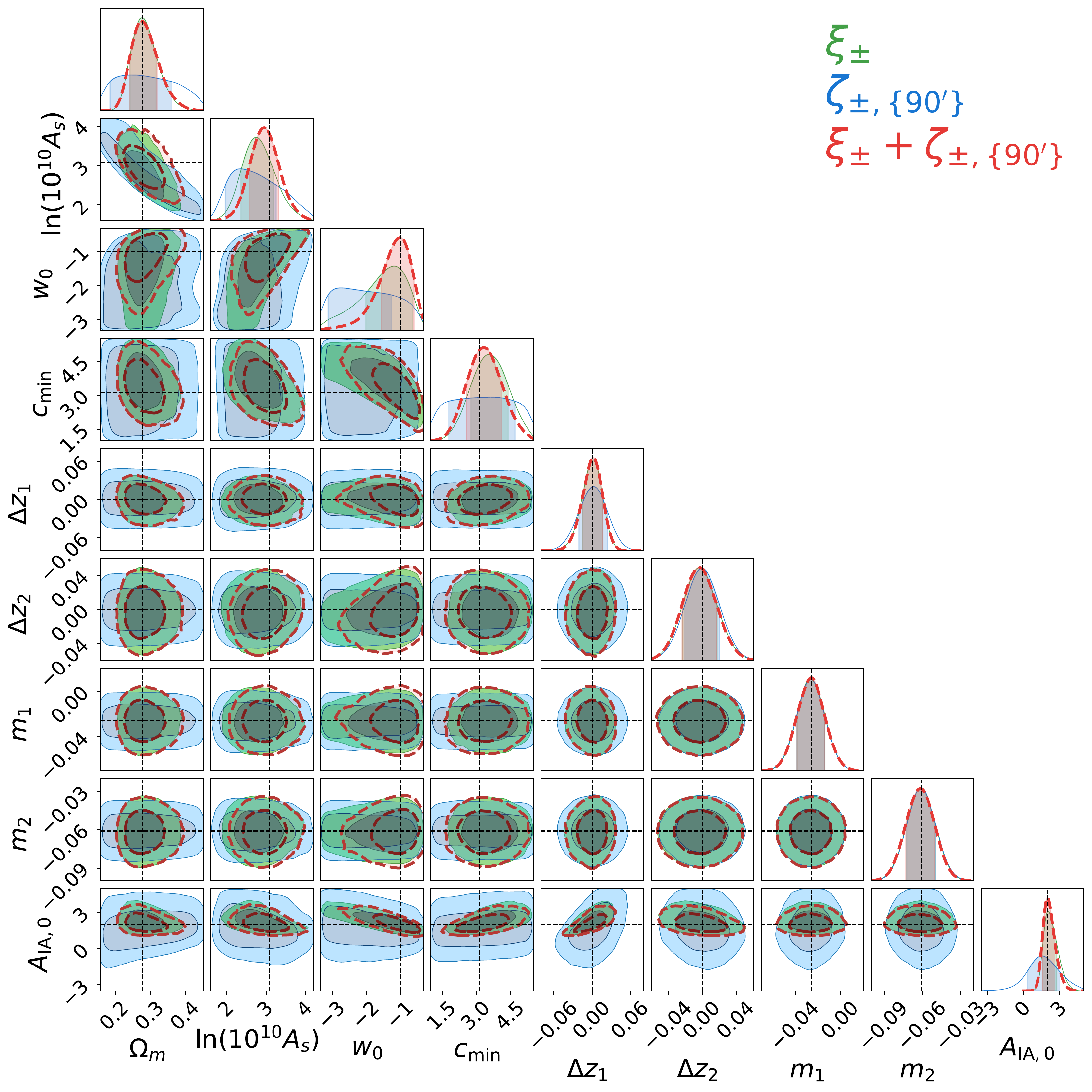}
    \caption{Parameter constraints obtained with our default NLA IA modelling, but on a data vector generated with the NLA IA model used in Refs.~\cite{2022PhRvD.106h3509G, 2021MNRAS.503.2300P, 2022MNRAS.516.1829P}. The result is for the T17 parameters (except we set $A_{\rm IA, 0} = 2$; cf.~dashed lines) and the FLASK covariance. The green, blue and red contours are for $\xi_{\pm}$, $\zeta_{\pm}$ and $\xi_{\pm} + \zeta_{\pm}$ constraints. We use DES Y3-like Gaussian priors for the systematic parameters.}
    \label{fig:NLA_bias}
\end{figure}

We turn our attention now to the impact of systematics (photo-$z$, shear calibration and IA) in $\zeta_{\pm}$ constraints. This is interesting as $\xi_{\pm}$ and $\zeta_{\pm}$ depend differently on systematics, and so combined analyses can potentially mitigate the degradation caused by these additional free parameters, leading to better cosmological constraints \cite{2006MNRAS.366..101H, 2008MNRAS.388..991S, 2012MNRAS.419.1804T}. Indeed, this has been studied recently in Ref.~\cite{2021MNRAS.503.2300P}, where it was shown that combining lensing 2- and 3-point correlation function information in a survey like Euclid could lead even to the self-calibration of the systematic parameters to levels that reduce the need for external calibration data sets.

The green and red contours in Fig.~\ref{fig:self_calibration} show the constraints for $\xi_{\pm}$ and $\xi_{\pm}+\zeta_{\pm}$, but instead of the tight DES Y3-like priors that we have assumed so far for the systematic parameters (cf.~Tab.~\ref{tab:param_prior}), we assume now wide priors for them. The result is for a noiseless realization of the data vector for the T17 parameters, with the exception that we set $A_{\rm IA, 0} = 2$ in this subsection. The improvements on the cosmological and baryonic parameters from adding $\zeta_{\pm}$ are $15.4\%$ for $\Omega_{\rm m}$, $8.8\%$ for ${\rm ln}\left(10^{10}A_s\right)$, $4.9\%$ for $w_0$ and $8.8\%$ for $c_{\rm min}$. Compared to the case where we marginalize over tight DES Y3-like Gaussian priors, varying the systematic parameters over wide priors degrades the improvement by factors of $2.3$, $7.8$ and $1.7$ for ${\rm ln}\left(10^{10}A_s\right)$, $w_0$ and $c_{\rm min}$ respectively. Furthermore, contrary to the case in Ref.~\cite{2021MNRAS.503.2300P}, the improvements that still exist do not appear to be associated with a significant self-calibration of the systematic parameters. This can be seen also on the left of Fig.~\ref{fig:self_calibration}, where the constraints on the systematic parameters in the combined $\xi_{\pm}+\zeta_{\pm}$ case (red) show improvements of $21\%$ for $\Delta z_1$, $8\%$ for $\Delta z_2$, $24\%$ for $m_1$, $17\%$ for $m_2$ and $18\%$ for $A_{\rm IA, 0}$. There is indeed a visible level of systematics self-calibration from combining $\xi_{\pm}$ with $\zeta_{\pm}$, but which still yields constraints that are substantially larger than using the externally calibrated DES Y3-like priors (blue).

The quantitative differences to the analysis of Ref.~\cite{2021MNRAS.503.2300P} could be at least partly due to some of the following reasons. First, Ref.~\cite{2021MNRAS.503.2300P} considers 3-point correlation function information by taking the equilateral lensing bispectrum as the data, whereas we consider $\zeta_{\pm}$ that probes predominantly the squeezed lensing bispectrum \cite{2021MNRAS.506.2780H, 2022MNRAS.515.4639H}. Second, Ref.~\cite{2021MNRAS.503.2300P} considers a treatment of the NLA IA model that is not the same as ours (cf.~App.~\ref{appendix:NLA}). Further, the results of Ref.~\cite{2021MNRAS.503.2300P} are based on Fisher matrix analyses, whereas ours are for simulated likelihood analyses with MCMC sampling. This can be especially important given how strongly non-Gaussian the marginalized posteriors of the systematic parameters are on the left of Fig.~\ref{fig:self_calibration}. Finally, our analysis is for a DES Y3-like survey assuming two tomographic bins, whereas Ref.~\cite{2021MNRAS.503.2300P} considers a larger Euclid-like survey with five tomographic bins, and thus a higher-dimensional subspace of systematic parameters. A deep investigation of the origin of the differences between the results of the two works would be interesting to pursue, but that is beyond the scope of the present paper.

We investigate also potential biases in the constraints of the $A_{\rm IA, 0}$ parameter from assuming different IA models in shear 3-point correlation function analyses. In particular, we wish to contrast the NLA model used in this paper (cf.~Sec.~\ref{sec:systematics} and App.~\ref{appendix:NLA}) with that in Ref.~\cite{2022PhRvD.106h3509G} which comes from Refs.~\cite{2021MNRAS.503.2300P, 2022MNRAS.516.1829P}. To do so we generate a noiseless data vector with the T17 parameters and $A_{\rm IA, 0} = 2$ {\it assuming the IA parametrization of Ref.~\cite{2022PhRvD.106h3509G}}, which we subsequently analyse by running MCMC constraints {\it assuming our IA modelling strategy}. At the $\xi_{\pm}$ level, the two IA treatments are equivalent, but there are differences at the level of the 3-point correlation functions (cf.~App.~\ref{appendix:NLA}).\footnote{Among other, the model of Ref.~\cite{2022PhRvD.106h3509G} includes terms $\propto A_{\rm IA, 0}^4$, whereas ours stops at third order $\propto A_{\rm IA, 0}^3$, as expected for a three-point correlation function.} Figure~\ref{fig:NLA_bias} shows the corresponding constraints for $\xi_{\pm}$ (green), $\zeta_{\pm}$ (blue) and $\xi_{\pm} + \zeta_{\pm}$ (red), with all yielding unbiased constraints, including $A_{\rm IA, 0}$. That is, at the level of the constraining power of our DES Y3-like setup, the differences between the two NLA IA models do not have any significant impact. We note, however, that whether the same conclusion holds for other survey setups should be checked on a case-by-case basis.

\subsection{The impact of different covariance estimates}
\label{sec:cov_impact_param_inference}

\begin{figure}
    \centering
    \begin{subfigure}{.5\textwidth}
        \centering
        \includegraphics[width=\linewidth]{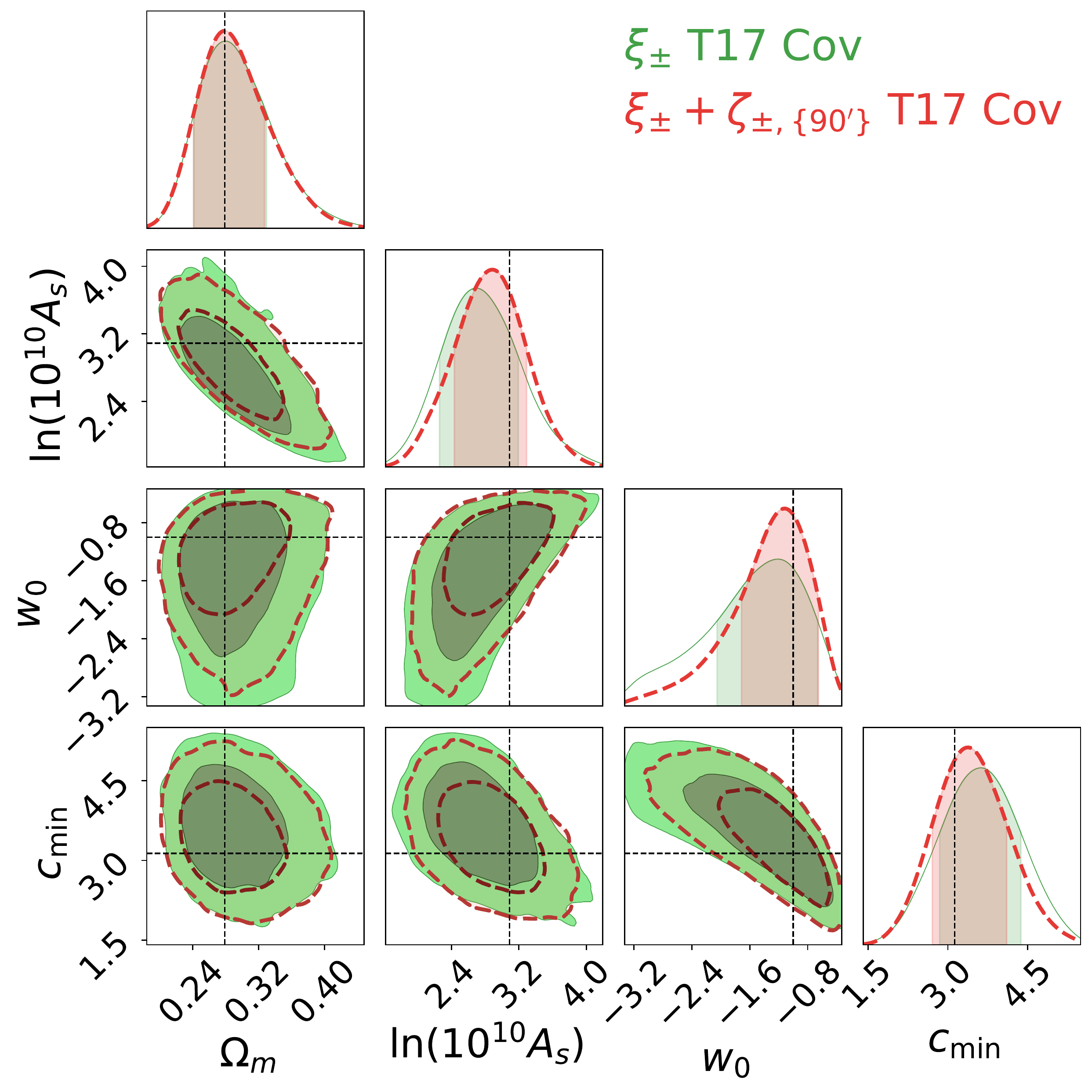}
        \label{fig:MCMC_T17_FLASK.a}
    \end{subfigure}%
    \begin{subfigure}{.5\textwidth}
        \centering
        \includegraphics[width=\linewidth]{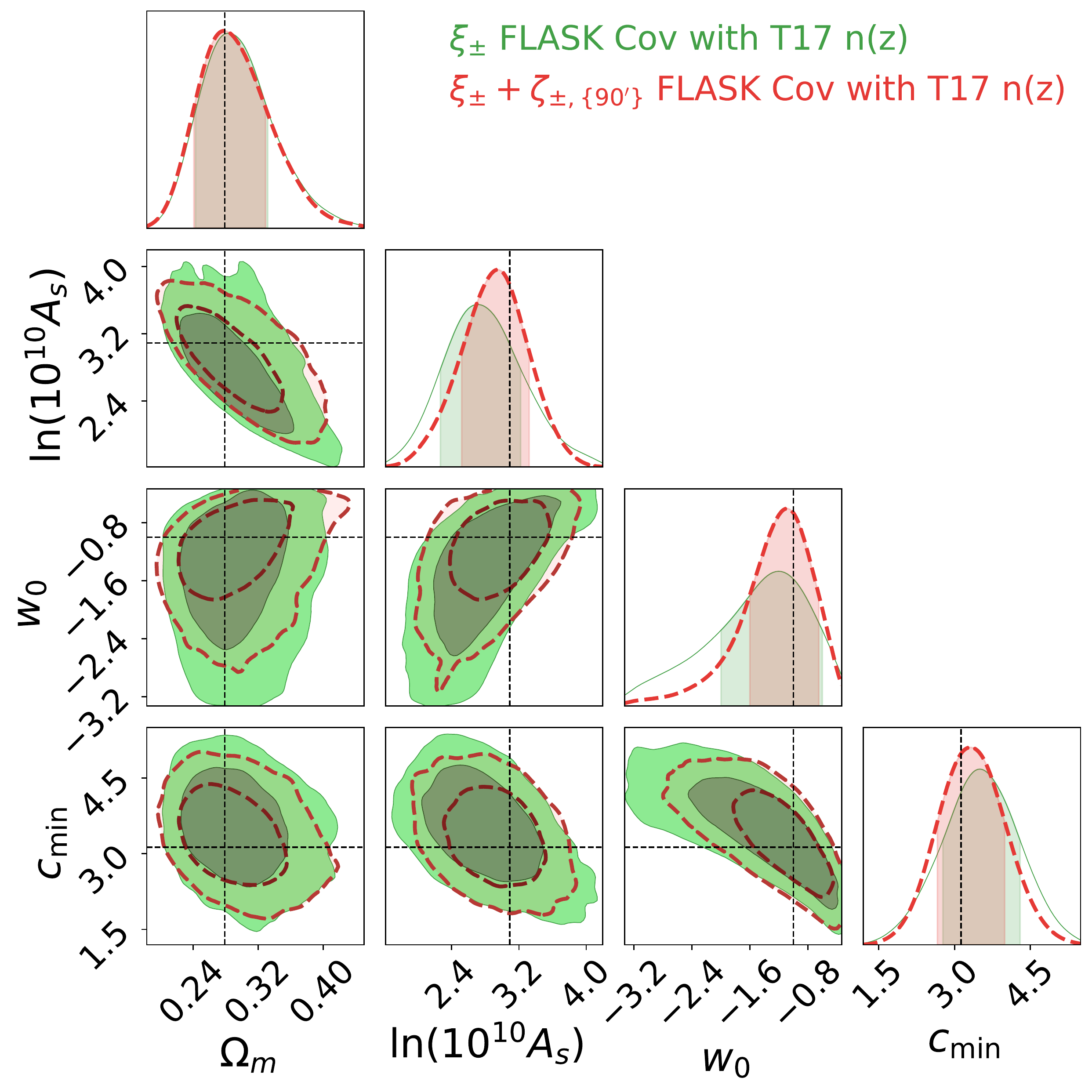}
        \label{fig:MCMC_T17_FLASK.b}
    \end{subfigure}
    \caption{Impact of the covariance matrix estimate on the parameter constraints. The left and right panels show the result for the covariance estimated from the T17 and FLASK shear maps, respectively. In both panels, the result is for a noiseless realization of the data vector from the theory model at the T17 parameters (dashed lines); the green and red contours are for $\xi_{\pm}$-only and $\xi_{\pm} + \zeta_{\pm}$, respectively. The result is for apertures with sizes $90\ {\rm arcmin}$ and systematics marginalized with the DES Y3-like Gaussian priors.}
    \label{fig:MCMC_T17_FLASK}
\end{figure}

We compare in Fig.~\ref{fig:MCMC_T17_FLASK} the parameter constraints obtained with the T17 covariance matrix (left) with those obtained using FLASK (right). In order to make a fair comparison, in this subsection we constructed a new FLASK covariance with the same number of footprint realizations as T17 ($N_s = 540$), and with a source redshift distribution matching the discretized one of the T17 simulations described in Sec.~\ref{sec:T17}. The result in Fig.~\ref{fig:MCMC_T17_FLASK} is for a noiseless realization of the data vector from the theory model at the T17 parameters, and with the systematic parameters marginalized with the DES Y3-like Gaussian priors. Table \ref{tab:MCMC_T17_FLASK} lists the corresponding improvements from adding $\zeta_{\pm}$ information to the constraints.

The two covariance matrices yield effectively the same parameter posteriors for $\xi_{\pm}$-only constraints; cf.~similarity between the green contours on the left and right of Fig.~\ref{fig:MCMC_T17_FLASK}. There are however some differences in the combined $\xi_{\pm} + \zeta_{\pm}$ constraints shown in red, with the FLASK covariance yielding smaller parameter error bars for most parameters. In particular, the improvements from $\zeta_{\pm}$ can be factors of $\approx 1.2-1.9$ larger with the FLASK covariance compared to T17.

The T17 and FLASK covariances in this subsection are estimated from ensembles of $540$ shear maps and so have the same noise level. These differences may indicate they are intrinsic to the different ability of $N$-body simulations and lognormal realizations to capture the covariance of $\zeta_{\pm}$,\footnote{The covariance of a 3-point function contains terms up to the 6-point function, which are not as faithfully captured in lognormal realizations, compared to $N$-body simulations.}, or due to residual statistical fluctuations for $N_s = 540$. We leave a more detailed investigation of the impact of the covariance matrix, including covariances calculated analytically \cite{2013MNRAS.429..344K, 2019JCAP...03..008B}, to future work.

\begin{table}
    \centering
    \begin{tabular}{|c|c|c|c|c|}
        \hline
        Covariance type & $\Omega_{\rm m}$ & ${\rm ln}\left(10^{10}A_s\right)$ & $w_0$ & $c_{\rm min}$ \\
        \hline
        \hline 
        FLASK (lognormal) & 1.1\% & 16.7\% & 32.1\% & 12.4\% \\
        \hline
        \hline
        T17 ($N$-body simulations) & 3.5\% & 8.8\% & 26.1\% & 8.7\% \\
        \hline
    \end{tabular}
    \caption{Impact of the covariance matrix estimate on the improvement of $\xi_{\pm} + \zeta_{\pm}$ constraints, relative to $\xi_{\pm}$-only. The result is for the aperture with size $90\ {\rm arcmin}$.}
    \label{tab:MCMC_T17_FLASK}
\end{table}
\section{Summary \& Conclusion}
\label{sec:conclusion}
The integrated shear 3PCF $\zeta_{\pm}$ \cite{2021MNRAS.506.2780H, 2022MNRAS.515.4639H} is a higher-order cosmic shear statistic that measures the correlation between the shear 2PCF measured in patches of the sky and the shear aperture mass in the same patches (cf.~Eq.~(\ref{eq:i3pcf_ensemble_average})). On small scales, $\zeta_{\pm}$ probes primarily the cosmological information encoded in the squeezed-limit lensing bispectrum. Two of the key advantages of $\zeta_{\pm}$ compared to other higher-order cosmic shear statistics are that (i) it can be straightforwardly evaluated from the data using efficient and well-tested 2-point correlation function estimators (i.e.~it does not explicitly require dedicated and more expensive 3-point estimators) and (ii) it admits a theoretical model based on the response approach to perturbation theory \cite{2022MNRAS.515.4639H} that is accurate in the nonlinear regime of structure formation, allowing to reliably account for the impact of baryonic physics. 

In this paper, we developed an analysis pipeline that can be directly applied to real cosmic shear data to obtain cosmological constraints from $\zeta_{\pm}$ and its combination with $\xi_{\pm}$. Compared to previous works on $\zeta_{\pm}$, the main significant advances in this paper are (i) the incorporation of lensing systematics associated with photo-$z$ uncertainties, shear calibration biases and galaxy IA (cf.~Sec.~\ref{sec:systematics}), and (ii) the development of a NN-based emulator for fast theory predictions to enable MCMC parameter inference. We tested our pipeline on a set of realistic cosmic shear maps based on $N$-body simulations, with DES Y3-like survey footprint, mask and source redshift distributions (cf.~Sec.~\ref{sec:cov}). 

In our tests of the analysis pipeline we have investigated in particular (i) the accuracy of the theory model (cf.~Sec.~\ref{sec:novel_bayesian}), (ii) the impact of the size of the apertures used to measure $\zeta_{\pm}$ (cf.~Sec.~\ref{sec:filter_optimisation}), (iii) the impact of lensing systematics (cf.~Sec.~\ref{sec:self_calibra_nla}) and (iv) the impact of $N$-body simulation vs.~lognormal estimates of the data vector covariance matrix (cf.~Sec.~\ref{sec:cov_impact_param_inference}). Our main findings can be summarized as follows:

\begin{itemize}
    \item Our analysis pipeline is accurate (cf.~Figs.~\ref{fig:xi_prediction_vs_t17} and \ref{fig:zeta_prediction_vs_t17}) and able to yield unbiased parameter constraints from our $N$-body simulation DES Y3-like data vectors (cf.~Fig.~\ref{fig:model_validation}).

    \item For the range of aperture sizes $\{50, 70, 90, 110, 130\}\ {\rm arcmin}$, $90\ {\rm arcmin}$ is what results in the largest information gain from $\zeta_{\pm}$. The combination of several filter sizes can improve the constraints further (cf.~Tab.~\ref{tab:single_and multiple_filter_param_improve}), but at the cost of dealing with a larger data vector and covariance matrix. 

    \item Although $\xi_{\pm}$ and $\zeta_{\pm}$ depend differently on the systematic parameters, we do not find significant improvements in their constraints in combined $\xi_{\pm} + \zeta_{\pm}$ analyses; i.e.~the mitigation of systematic effects still requires prior calibration from external data (cf.~Fig.~\ref{fig:self_calibration}). This is in contrast with the findings in Ref.~\cite{2021MNRAS.503.2300P}, although this may be due to differences in the 3-point correlation function studied, survey setup and other analysis details. At the level of the DES Y3 constraining power, different modelling strategies for IA lead also to no significant biases in parameter constraints (cf.~Fig.~\ref{fig:NLA_bias}).

    \item Relative to $\xi_{\pm}$-only constraints with the $N$-body covariance matrix, adding $\zeta_{\pm}$ leads to improvements of $4\%$ for $\Omega_{\rm m}$, $9\%$ for ${\rm ln}\left(10^{10}A_s\right)$, $26\%$ for $w_0$ and $9\%$ for $c_{\rm min}$. Except for $\Omega_{\rm m}$, these are factors of $\approx 1.2-1.9$ smaller compared to the FLASK covariance. This may be due to residual statistical fluctuations at the level of our number of simulation realizations ($N_s = 540$), or simply that lognormal realizations do not provide reliable estimates of the $\zeta_{\pm}$ covariance matrix.
\end{itemize}

Overall, our results corroborate with a realistic MCMC-based simulated likelihood analysis the encouraging findings from previous idealized Fisher matrix forecasts \cite{2021MNRAS.506.2780H, 2022MNRAS.515.4639H}. The analysis pipeline developed and tested here can be readily applied to real survey data, enabling the exploration of the potential of the integrated shear 3PCF $\zeta_{\pm}$ to improve cosmological parameter constraints using cosmic shear observations.

\acknowledgments
We would like to thank Pierre Burger, Juan M. Cruz-Martinez, Chris Davies, Mariia Gladkova, Eiichiro Komatsu, Elisabeth Krause and Alessio Spurio-Mancini for very helpful comments and discussions at various stages of this project. We acknowledge support from the Excellence Cluster ORIGINS which is funded by the Deutsche Forschungsgemeinschaft (DFG, German Research Foundation) under Germany’s Excellence Strategy - EXC-2094-390783311. Most of the numerical calculations have been carried out on the ORIGINS computing facilities of the Computational Center for Particle and Astrophysics (C2PAP). We would like to particularly thank Anthony Hartin for the support in accessing these computing facilities. The results in this paper have been derived using the following publicly available libraries and software packages: \texttt{healpy} \cite{2019JOSS....4.1298Z}, \texttt{Treecorr} \cite{2004MNRAS.352..338J}, \texttt{CLASS} \cite{2011JCAP...07..034B}, \texttt{FLASK} \cite{2016MNRAS.459.3693X}, 
\texttt{Vegas}
\cite{2021JCoPh.43910386L},
\texttt{Cosmopower} \cite{2022MNRAS.511.1771S}, \texttt{GPflow} \cite{2016arXiv161008733M} and \texttt{Numpy} \cite{2020Natur.585..357H}. We also acknowledge the use
of \texttt{matplotlib} \cite{2007CSE.....9...90H} and \texttt{ChainConsumer} \cite{2016JOSS....1...45H} python packages in producing the figures shown in this paper.

\paragraph{Data availability}

The numerical data underlying the analysis of this paper may be shared upon reasonable request to the authors. 

\appendix
\section{The modelling of intrinsic alignments}
\label{appendix:NLA}

In this appendix we describe our modelling of galaxy intrinsic alignments in $\xi_{\pm}$ and $\zeta_{\pm}$.


\subsection*{General considerations}

The observed galaxy ellipticity in cosmic shear observations $\eps_{\rm obs}$ is a combination of the gravitational (G) lensing shear component $\gamma$ and the intrinsic (I) ellipticity of the galaxies $\eps_{\rm I}$ induced by correlations with local gravitational tidal fields at the source (in this appendix, we ignore the random stochastic component that would contribute as shape noise):
\bq\label{eq:IA}
\eps^{i}_{\rm obs}(\vtheta) = \gamma^{i}(\vtheta) + \eps^{i}_{\rm I}(\vtheta),
\eq
where $i$ denotes a specific source galaxy redshift bin. The lensing shear is related to the lensing convergence $\kappa$ as \cite{2006glsw.conf.....M}
\bq\label{eq:gammakappa}
\gamma(\vell) = e^{2i\phi_\ell} \kappa(\vell)\ \ ;\ \ \kappa(\vell) = \int {\rm d}^2\vtheta \ \kappa(\vtheta)e^{-i\vell\vtheta}\ \ ; \ \ \kappa(\vtheta) = \int {\rm d}\chi \ q(\chi)\delta_m(\vtheta\chi, \chi),
\eq
where $\delta_m$ is the three-dimensional matter density contrast.\footnote{To ease the notation, we distinguish between real- and harmonic-space variables by their arguments. For example, $\kappa(\vtheta)$ and $\kappa(\vell)$ are the lensing convergence in real and harmonic space, respectively.} In analogy, we can write for the intrinsic component $\eps_{\rm I}$
\bq\label{eq:epsIkappa}
\eps_{\rm I}(\vell) = e^{2i\phi_\ell} \kappa_{\rm I}(\vell)\ \ ;\ \ \kappa_{\rm I}(\vell) = \int {\rm d}^2\vtheta \ \kappa_{\rm I}(\vtheta)e^{-i\vell\vtheta}\ \ ; \ \ \kappa_{\rm I}(\vtheta) = \int {\rm d}\chi \ n(\chi)\delta_{\rm I}(\vtheta\chi, \chi),
\eq
where $\delta_{\rm I}$ is a three-dimensional field that determines {\it effectively} the intrinsic alignment (IA) of the galaxies with their local gravitational tidal fields; note also that the line-of-sight kernel is now just the source galaxy distribution $n(\chi)$, and not the lensing kernel $q(\chi)$.

In the popular {\it nonlinear linear alignment} (NLA) model \cite{2007MNRAS.381.1197H, 2007NJPh....9..444B}, one writes 
\bq\label{eq:NLAdef}
\delta_{\rm I}(\vx, z) = f_{\rm IA}(z)\delta_m (\vx, z),
\eq
treating $\delta_m$ as the nonlinear matter density contrast. The amplitude $f_{\rm IA}(z)$ is
\bq\label{eq:fia}
f_{\text{IA}}(z) = -A_{\text{IA},0}\left(\frac{1+z}{1+z_0}\right)^{\alpha_{\text{IA}}}\frac{c_1\rho_{\text{crit}}\Omega_{\text{m},0}}{D(z)} \ ,
\eq
where $A_{{\rm IA},0}$, $\alpha_{\rm IA}$ are free redshift-independent parameters, $c_1 = 5\times10^{-14}$  $(h^2 M_{\odot}/{\rm Mpc^3})^{-1}$ \cite{2007NJPh....9..444B}, $\rho_{\rm crit}$ is the critical cosmic energy density, $D(z)$ is the growth factor normalized to unity today, and $z_0$ is some reasonable pivot redshift value. 

Note that this is only an {\it effective} parametrization of the impact of IA in cosmic shear observations. A more rigorous approach would involve a description of the relation of galaxy shapes and tidal fields in 3D, subsequently projected to the sky plane. This is the approach described in Refs.~\cite{2020JCAP...01..025V, 2021JCAP...05..061V} based on bias expansions in effective field theory, which is however valid only in the quasi-linear, large-scale regime of structure formation. Extensions of the NLA model to include nonlinear corrections to Eq.~(\ref{eq:NLAdef}) also exist \cite{2019PhRvD.100j3506B}.


\subsection*{Contributions to $\xi_{\pm}$}

The two shear 2PCF $\xi_{\pm}$ are given by 
\bq\label{eq:xipm}
\xi^{ij}_{+,{\rm obs}}(\alpha) &=& \langle  \eps^i_{\rm obs}(\vtheta) \eps^{j*}_{\rm obs}(\vtheta + \valpha) \rangle \\
\xi^{ij}_{-,{\rm obs}}(\alpha) &=&  \langle \eps^i_{\rm obs}(\vtheta) \eps^j_{\rm obs}(\vtheta + \valpha) e^{-4i\phi_\alpha} \rangle,
\eq
and each can be decomposed into GG, GI, IG and II terms as
\bq
\xi^{ij}_{\pm,{\rm obs}} = \xi^{ij}_{\pm,{\rm GG}} + \xi^{ij}_{\pm,{\rm GI}} + 
\xi^{ij}_{\pm,{\rm IG}} +
\xi^{ij}_{\pm,{\rm II}}. 
\eq
The GI case of $\xi^{ij}_{+, \rm obs}$, for example, is given by (the derivations are analogous for all terms):
\bq\label{eq:xipmIG}
\xi^{ij}_{+, {\rm GI}}(\alpha) &=& \langle  \gamma^i(\vtheta) \eps^{j*}_{\rm I}(\vtheta + \valpha) \rangle \nonumber \\
&=& \int \frac{{\rm d}\ell\ell}{2\pi}P^{ij}_{\kappa\kappa_{\rm I}}(\ell)J_0(\ell\alpha),
\eq
where $P^{ij}_{\kappa\kappa_{\rm I}}(\ell)$ is defined as $(2\pi)^2 P^{ij}_{\kappa\kappa_{\rm I}}(\ell)\delta_D(\vell + \vell') = \langle\kappa^i(\vell)\kappa^j_{\rm I}(\vell')\rangle$ and given by
\bq
P^{ij}_{\kappa\kappa_{\rm I}}(\ell) = \int {\rm d\chi} \frac{q^{i}(\chi) n^{j}(\chi)}{\chi^2} P^{\rm 3D}_{\delta_{m}\delta_{\rm I}}(\ell/\chi, \chi).
\eq
The $P^{\rm 3D}_{\delta_{m}\delta_{\rm I}}$ is defined as $(2\pi)^3 P^{\rm 3D}_{\delta_{m}\delta_{\rm I}} (\vk_1 + \vk_2) = \langle\delta_m(\vk_1)\delta_{\rm I}(\vk_2)\rangle$, and in the NLA model it is 
\bq
P^{\rm 3D}_{\delta_m\delta_{\rm I}}(k,z) = f_{\rm IA}(z) P^{\rm 3D}_{\delta_m\delta_m}(k,z).
\eq
That is, the GI contribution to $\xi^{ij}_{+, \rm obs}$ can be obtained by replacing the $j$th lensing kernel $q^j(\chi)$ in the expression of the GG term with $n^j(\chi) f_{\rm IA}$. It follows as a result that all contributions from GG, GI, IG and II can be obtained by replacing all lensing kernels $q(\chi)$ with $q(\chi) + n(\chi) f_{\rm IA}$, as in Eq.~(\ref{eq:lensing_kernel_q_with_NLA}). This yields terms $\propto f_{\rm IA}^0$ (GG), $\propto f_{\rm IA}$ (GI, IG) and $\propto f_{\rm IA}^2$ (II).


\subsection*{Contributions to $\zeta_{\pm}$}

The observed integrated shear 3PCF $\zeta_{\pm}$ is defined as
\bq\label{eq:zetadef}
\zeta^{ijk}_{\pm, \rm obs}(\alpha) = \Big< \hat{M}^{i}_{\rm ap, obs}(\vtheta_C) \hat{\xi}^{jk}_{\pm, {\rm obs}}(\alpha; \vtheta_C) \Big>.
\eq
The position-dependent shear 2PCF $\hat{\xi}^{jk}_{\pm, {\rm obs}}(\alpha; \vtheta_C)$ also contains GG, GI, IG and II terms. Further, the IA terms also contribute to the 1-point aperture mass $\hat{M}^i_{\rm ap, obs}(\vtheta_C)$, which contains G and I terms as
\bq\label{eq:MapnoIA}
\hat{M}^i_{\rm ap, obs}(\vtheta_C) &=& \int {\rm d}^2\vtheta \Big[\kappa^i(\vtheta) + \kappa^i_{\rm I}(\vtheta)\Big]U(\vtheta_C - \vtheta).
\eq
This thus generates the following 8 contributions to $\zeta^{ijk}_{\pm, \rm obs}(\alpha)$:
\bq
\zeta^{ijk}_{\pm, \rm obs} = \zeta^{ijk}_{\pm, \rm GGG} + \zeta^{ijk}_{\pm, \rm GGI} + 
\zeta^{ijk}_{\pm, \rm GIG} +
\zeta^{ijk}_{\pm, \rm GII} + 
\zeta^{ijk}_{\pm, \rm IGG} + 
\zeta^{ijk}_{\pm, \rm IGI} + 
\zeta^{ijk}_{\pm, \rm IIG} + 
\zeta^{ijk}_{\pm, \rm III}.
\eq
Again, just as a single example, the IIG case for $\zeta_{+, \rm obs}$ can be written as
\bq\label{eq:zetaIIG}
\zeta^{ijk}_{+, \rm IIG}(\alpha) = \frac{1}{A(\alpha)} \int \frac{{\rm d}\ell \ell}{2\pi}\mathcal{B}^{ijk}_{+, \rm IIG}(\ell) J_{0}(\ell\alpha),
\eq
where
\bq\label{eq:BkappaIIG}
\mathcal{B}^{ijk}_{+, \rm IIG}(\ell) &=& \frac{\text{d}^2\boldsymbol{\ell}_1}{(2\pi)^2} \int\frac{\text{d}^2\boldsymbol{\ell}_2}{(2\pi)^2} B^{ijk}_{\kappa_{\rm I}\kappa_{\rm I}\kappa}\left(\vell_1, \vell_2, -\vell_{12}\right) e^{2i\left(\phi_2 - \phi_{-1-2}\right)} U(\vell_1)W(\vell+\vell_2)W(-\vell-\vell_{12}), \nonumber \\
\eq
with $\vell_{12} = \boldsymbol{\ell}_1 + \boldsymbol{\ell}_2$ and
\bq
B^{ijk}_{\kappa_{\rm I}\kappa_{\rm I}\kappa_{\rm G}}\left(\vell_a, \vell_b, \vell_c\right) = \int {\rm d}\chi \frac{n^{i}(\chi)n^{j}(\chi)q^{k}(\chi)}{\chi^4} B^{\rm 3D}_{\delta_{\rm I}\delta_{\rm I}\delta_m}\left(\frac{\vell_a}{\chi}, \frac{\vell_b}{\chi}, \frac{\vell_{c}}{\chi}; \chi\right).
\eq
The derivation of these expressions is the same as the usual gravitational lensing GGG expression, except one replaces the first two instances of $\kappa$ by $\kappa_{\rm I}$. In the NLA model, $B^{\rm 3D}_{\delta_{\rm I}\delta_{\rm I}\delta_m} =f_{\rm IA}^2 B^{\rm 3D}_{\delta_{m}\delta_{m}\delta_m}$. That is, the IIG contribution to $\zeta^{ijk}_{+, \rm obs}(\alpha)$ term can be obtained from GGG by simply replacing the $i$th and $j$th lensing kernels $q^{i}(\chi)$, $q^{j}(\chi)$ with $n^{i}(\chi) f_{\rm IA}$ and $n^{j}(\chi) f_{\rm IA}$. It follows as a result that all of the 8 contributions to $\zeta^{ijk}_{\pm, \rm obs}(\alpha)$ can be obtained by replacing all lensing kernels $q(\chi)$ with $q(\chi) + n(\chi) f_{\rm IA}$, as in Eq.~(\ref{eq:lensing_kernel_q_with_NLA}). This yields terms $\propto f_{\rm IA}^0$ (GGG), $\propto f_{\rm IA}$ (GGI, GIG, IGG) and $\propto f_{\rm IA}^2$ (GII, IGI, IIG) and $\propto f_{\rm IA}^3$ (III).

These 3-point contributions from galaxy IA are different than those derived in Ref.~\cite{2021MNRAS.503.2300P} using also the NLA model. Among other differences, their III term is $\propto f_{\rm IA}^4$ and their GII + IGI + IIG terms are $\propto f_{\rm IA}^3$ (cf.~their Eqs.~(30 - 32)). Reference~\cite{2021MNRAS.503.2300P} does not provide a detailed derivation of their expressions, which keeps us from inspecting this issue further. We emphasise, however, that the NLA model is in itself only an approximation of the effect of galaxy IA on small-scales, and so even our expressions should be interpreted in light of this.


\bibliographystyle{utphys}
\bibliography{refs}

\end{document}